\PassOptionsToPackage{dvipsnames}{xcolor}
\documentclass[sn-mathphys,Numbered]{sn-jnl}%

\usepackage{geometry}
\usepackage{subcaption}
\usepackage{graphicx}%
\usepackage{amsmath,amssymb,amsfonts}%
\usepackage{amsthm}%
\usepackage{mathrsfs}%
\usepackage[title]{appendix}%
\usepackage{xcolor}%
\usepackage{multicol}%
\usepackage{multirow}%
\usepackage{tabularray}%
\usepackage{textcomp}%
\usepackage{manyfoot}%
\usepackage{booktabs}%
\usepackage[ruled]{algorithm2e}
\usepackage{algorithmicx}%
\usepackage{algorithm2e}
\usepackage{float}

\SetKwInput{kwInit}{Init}

\usepackage{algpseudocode}%
\usepackage{listings}%

\theoremstyle{thmstyleone}%

\theoremstyle{thmstyletwo}%

\theoremstyle{thmstylethree}%

\begin{document}

\title[Article Title]{CAMP: Continuous and Adaptive Learning Model in Pathology }%

\author[a]{\fnm{Anh Tien Nguyen}}\email{ngtienanh@korea.ac.kr}
\author[a]{\fnm{Keunho Byeon}}\email{bkh5922@korea.ac.kr}
\author[b,c]{\fnm{Kyungeun Kim}}\email{kekim23@naver.com}
\author[b]{\fnm{Boram Song}}\email{setmefri62@gmail.com}
\author[b]{\fnm{Seoung Wan Chae}}\email{chae\_sw@hanmail.net}
\author*[a]{\fnm{Jin Tae Kwak}}\email{jkwak@korea.ac.kr}

\affil[a]{\orgdiv{School of Electrical Engineering}, \orgname{Korea University}, \orgaddress{\street{Anam-ro}, \city{Seoul}, \postcode{02841}, \country{South Korea}}}
\affil[b]{\orgdiv{Department of Pathology, Kangbuk Samsung Hospital}, \orgname{Sungkyunkwan University School of Medicine}, \orgaddress{\street{Saemunan-ro}, \city{Seoul}, \postcode{03181}, \country{South Korea}}}
\affil[c]{\orgdiv{Pathology Center}, \orgname{Seegene Medical Foundation}, \orgaddress{\street{Cheonho-daero}, \city{Seoul}, \postcode{133847}, \country{South Korea}}}

\abstract{
    There exist numerous diagnostic tasks in pathology. Conventional computational pathology formulates and tackles them as independent and individual image classification problems, thereby resulting in computational inefficiency and high costs. 
    To address the challenges, we propose a generic, unified, and universal framework, called a continuous and adaptive learning model in pathology (CAMP), for pathology image classification. CAMP is a generative, efficient, and adaptive classification model that can continuously adapt to any classification task by leveraging pathology-specific prior knowledge and learning task-specific knowledge with minimal computational cost and without forgetting the knowledge from the existing tasks. 
    We evaluated CAMP on 22 datasets, including 1,171,526  patches and 11,811 pathology slides, across 17 classification tasks. 
    CAMP achieves state-of-the-art classification performance on a wide range of datasets and tasks at both patch- and slide-levels and reduces up to 94\% of computation time and 85\% of storage memory in comparison to the conventional classification models.
    Our results demonstrate that CAMP can offer a fundamental transformation in pathology image classification, paving the way for the fully digitized and computerized pathology practice.
    }
    

\maketitle

\section{Introduction}\label{sec1}

    With the rapid advances in artificial intelligence (AI) and imaging techniques and easy access to digital systems, computational pathology is promising to revolutionize and evolve the pathology landscape at an unprecedented pace \cite{Cui2021}. A recent study demonstrates that the impact of computational pathology will be significant in many aspects of the pathology workflow \cite{alvaro}, including but not limited to disease detection and diagnosis (e.g., lymphovascular invasion detection \cite{CHEN202326,Bejnordi} and colorectal cancer grading \cite{LEE2023107749,joint,camel}), quantification (e.g., counting nuclei \cite{graham2021conic,chen_count_nuclei} or mitosis \cite{Cree2021,Tabata2019} and quantification of biomarkers \cite{STALHAMMAR2016318,Vlad_biomaker}), standardization of the slide preparation \cite{Yagi2011,Barisoni_standardization}, and quality control and assurance of whole slide images and reports \cite{Chen_wsi_assess,Avanaki_quality_control,Ameisen2014}. However, a limited number of computational pathology tools have been adopted as a part of the routine clinical workflow \cite{steiner}. Therefore, gaps or barriers exist in translating computational pathology tools into clinical practice.

    A large portion of routine pathology practice can be formulated as an image classification task where an examiner (i.e., a pathologist) assigns a class label to an image of interest (e.g., biopsy specimens). Class labels can vary from the presence of cancer and metastasis, histological sub-types, to survival rate of subjects. 
    To tackle such classification tasks, the current practice of computational pathology, in general, focuses on a single task at a time such that an individual and independent AI model, built based upon convolutional neural networks (CNNs) \cite{joint,camel,cnn_copath,cnn_copath_2,HEKLER201991} and/or vision transformers (ViT) \cite{ctranspath, ordervit,vit_copath}, is developed and validated per classification task. This approach has two major drawbacks. 
    First, it cannot fully utilize the existing knowledge and resources. The characteristics of tissues among different tasks can be shared. For example, there can be two tasks for colorectal tissues such as colorectal cancer grading with 4 categories (\textit{benign}, \textit{well differentiated cancer}, \textit{moderately differentiated cancer}, and \textit{poorly differentiated cancer}) and colorectal tissue sub-typing with 7 categories (\textit{adipose}, \textit{background}, \textit{debris}, \textit{lymphocyte}, \textit{normal}, \textit{stroma}, and \textit{tumor}). The structure and shape of liver cancers (\textit{benign}, \textit{grade 1}, \textit{grade 2}, and \textit{grade 3}) and kidney cancers (\textit{benign}, \textit{grade 1}, \textit{grade 2}, \textit{grade 3}, and \textit{grade 4}) are analogous to each other. As AI models are individually and independently developed and validated, taking advantage of other related tissues and tasks is challenging.
    Second, it is not scalable. Some showed that the same AI model can be adopted for other classification tasks \cite{ctranspath}, but one still needs to repeat the entire training and validation process per task. Though successful in resolving each task, this approach inevitably results in numerous computational pathology tools, as many as classification tasks in pathology, to be implemented and utilized in clinics. Consequently, this comes at the cost of computational resources, maintenance, and energy. The more tools we use in clinics, the higher the cost and complexity it may add up. Neither scientific nor medical communities have taken such costs and issues into account.

    There are two ways to tackle the above problem. The first approach is to develop a unified AI model for all classification tasks \cite{GPC}. As a new task is incorporated, the previous universal AI models need full training and validation for the new and existing tasks. Due to the vast number of tasks and data samples per task, training and deployment require a tremendous amount of time, which considerably limits the applicability of the method, and thus, it is infeasible. 
    The second approach is to utilize a so-called foundation model that can be applied to a wide range of applications \cite{plip,ctranspath,uni,quilt1m,phikon}. These models have recently drawn significant attention for their superior learning capability \cite{quilt1m}. These can be applied to differing classification tasks with and without adaptation procedures via zero-shot learning. 
    The most common adaptation strategy is fine-tuning and/or linear probing, yet no optimal adaptation strategy for each task is available. The more fine-tuned or adjusted the model is, the higher performance it achieves per task. However, this approach suffers from catastrophic forgetting, which is a phenomenon where AI models lose the information from the previous tasks as learning or adapting to a new task \cite{french1999catastrophic,kirkpatrick2017overcoming}. Using foundation models without adaptation is also not an option since there is a considerable performance gap between zero-shot learning and traditional supervised learning.
    Therefore, the field of computational pathology needs not only a new type of AI model but also an efficient and effective manner of training and adaptation methodologies to handle a variety of classification tasks together without substantial loss of information and performance.

    In this study, we propose a Continuous and Adaptive learning Model in Pathology, so-called CAMP, as a generic, unified, and universal framework for pathology image classification, which addresses the challenges and limitations of the current pathology image classification approaches in computational pathology as outlined above.
    The major strength of CAMP is four-fold. First, CAMP is a generative model. It transforms or reformulates the image classification problems as text generation problems; for instance, given a pathology image, CAMP directly generates a text phrase or label such as \textit{in situ carcinoma} and \textit{mucus} instead of choosing an index designated to the particular class label. 
    Second, CAMP is adaptive. It can adapt to a given classification task without losing prior knowledge and classification performance, allowing it to learn from new tasks continuously.  
    Third, CAMP is efficient. To adapt to a new task, CAMP only trains an minor number of learnable parameters for new task-specific knowledge, while the common knowledge and other tasks' knowledge are decoupled and preserved. Therefore, minimal modifications and costs are required for the adaptation, which maintains the efficiency of CAMP when increasing the number of downstream tasks.
    Fourth, CAMP is versatile. It is able to conduct various classification tasks in pathology at both patch level and whole slide image (WSI) level with high accuracy by adapting itself to each task efficiently and effectively. 
    In experiments with 17 classification tasks, including 1,171,526 patches and 11,811 slides from 22 pathology image classification datasets originating from 8 different organs (Fig. \ref{fig_dataset}), we demonstrate that CAMP is highly adaptive and efficient in learning and conducting a variety of classification tasks as well as can achieve highly accurate classification results regardless of the types of classification tasks, organs, and datasets.

    We build CAMP under the following hypotheses: 1) there exists common knowledge for pathology image analysis that is applicable to any classification tasks; 2) tasks are distinctive from each other, and thus, there also exists task-specific knowledge; 3) both common and task-specific knowledge is required to achieve high performance in each task.
    In order to utilize the common knowledge, we adopt the pre-trained weights trained on a large-scale pathology image dataset. As for the task-specific knowledge, we employ adaptors that are adjusted and optimized per task. Then, the weights of the adaptors (i.e., task-specific knowledge) are added to the pre-trained weights (i.e., common knowledge on pathology images) to conduct a particular classification task.

    CAMP receives two inputs, including a pathology patch/slide and a text prompt.
    The text prompt instructs CAMP to which task it needs to conduct, such as \textit{``The cancer sub-type of this breast tissue is''}. 
    CAMP processes the two inputs and generates the text label by combining and utilizing both a visual model (a visual encoder) and a language model (a text decoder). 
    The visual model extracts image features from the input pathology image. The image features are fused with the text embedding, extracted from the text prompt by the language model, and fed into the language model to produce the text label in an auto-regressive manner. In CAMP, common knowledge in a single visual and language model is sufficient to perform numerous classification tasks. In other words, the same visual and language model is shared among various types of classification tasks. 
    The conventional methods, however, need to adopt at least two separate layers with the same or differing number of neurons (or processing units) to conduct two classification tasks together. Though the intermediate layers can be shared between two and employed from the previous models via transfer learning, the new layers are often randomly initialized. These may contribute to the increase in the size and complexity of the classification models and the decrease in the classification performance due to the lack of prior knowledge of pathology. However, CAMP does not suffer from such issues with computational complexity and performance degradation, holding the potential for transforming the approaches to classification tasks.

    CAMP is a paradigm shift for image classification tasks in computational pathology, transitioning from the long-lasting discriminating approaches to the generative approaches, from the category assignment to the text generation, and from static learning to dynamic and continual learning (Fig. \ref{fig_patch_workflow} and \ref{fig_slide_workflow}). 
    We systematically evaluate the ability of CAMP on one of the most extensive collections of pathology images and tasks ever used together for image classification tasks (Fig. \ref{fig_dataset}).
    We show that CAMP is superior to the conventional image classification models in computational pathology and other domains. We also investigate the effect of the prior knowledge, i.e., pre-trained weights and the text prompt, on the classification performance. Moreover, we examine the computational requirements of CAMP and other methods to validate the scalability and utility of CAMP in clinics.

\section{Method}
    \subsection{CAMP}

\begin{figure}[!ht]%
            \centering
            \includegraphics[width=1\textwidth]{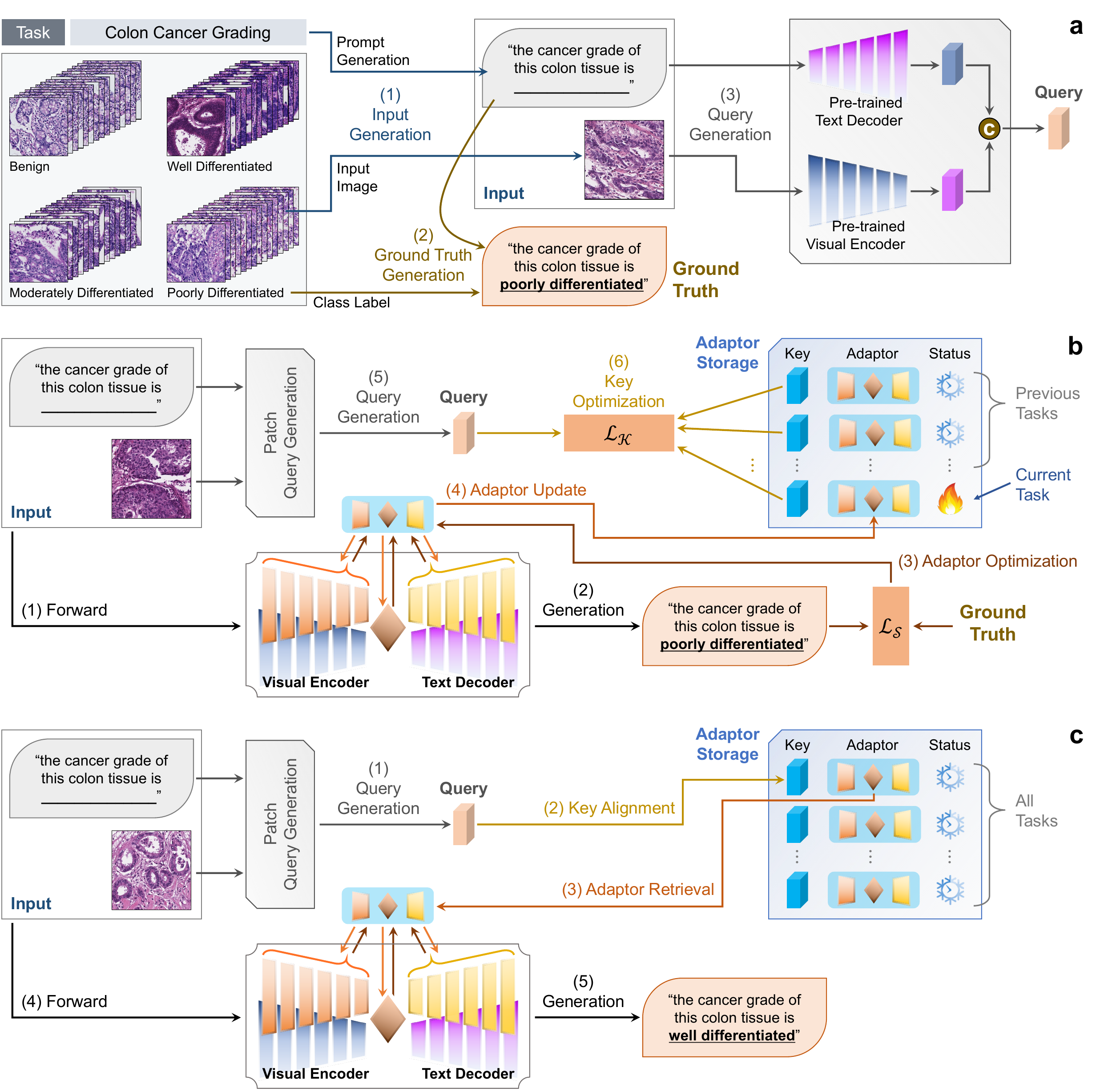}
            \caption{Overview of CAMP for \textbf{patch-level classification}. $\bold{a}$) For each patch classification task, the image-text prompt input and text ground truth are generated. The patch query generation is generated by a pre-trained visual encoder and a pre-trained text decoder. $\bold{b}$) During training, $\mathcal{L}_{\mathcal{S}}$ is used for optimizing adaptors, whereas $\mathcal{L}_{\mathcal{K}}$ is utilized for updating a key. This process only updates the training task and preserves the knowledge of previously learned tasks. $\bold{c}$) During inference, a query is generated based on an input to retrieve the most suitable adaptors. After being integrated with the adaptors, CAMP generates a textual prediction.}\label{fig_patch_workflow}
        \end{figure}

\begin{figure}[!ht]%
            \centering
            \includegraphics[width=1\textwidth]{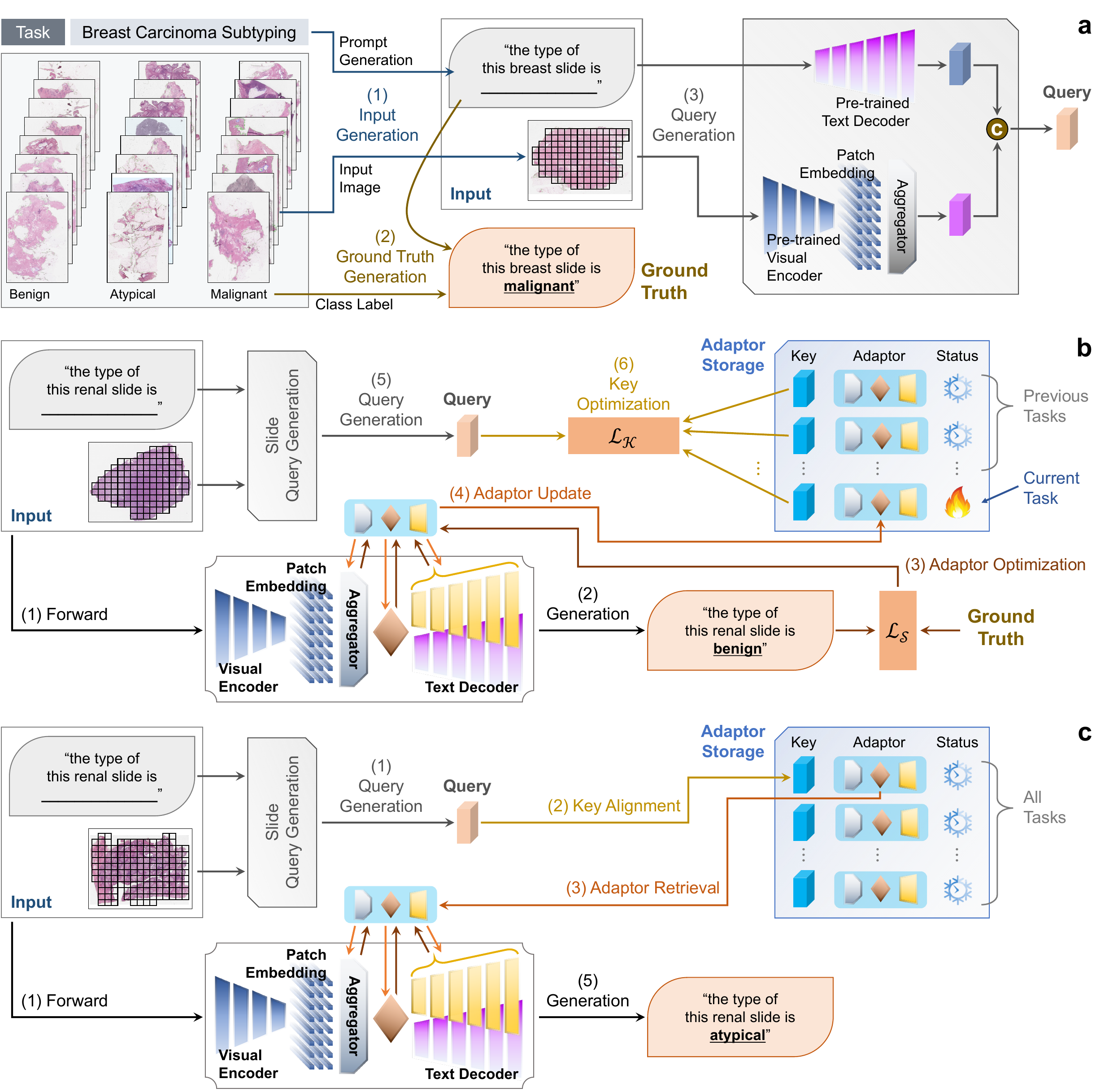}
            \caption{Overview of CAMP for \textbf{slide-level classification}. $\bold{a}$) For each slide classification task, the image-text prompt input and text ground truth are generated. The slide query generation is produced by a pre-trained visual encoder, a pre-trained text decoder, and a non-parametric aggregator. $\bold{b}$) Similar to patch-level, $\mathcal{L}_{\mathcal{S}}$ and $\mathcal{L}_{\mathcal{K}}$ are used for optimizing adaptors and a key during training a current task. A visual encoder is frozen in this process. $\bold{c}$) The slide-level inference is similar to patch-level, except for the adaptors. Note that the aggregator (blue) in the generative model is parametric, which is different from the non-parametric aggregator (grey) in the query generation procedure.} \label{fig_slide_workflow}
        \end{figure}

        CAMP is a highly efficient and easily adaptable framework for patch- and whole-slide-level image classification tasks in computational pathology. The framework consists of three primary components: 1) a visual encoder $\mathcal{V}$, 2) a text decoder $\mathcal{T}$, and 3) an adaptor storage $\mathcal{S}$. 
        $\mathcal{V}$ receives a pathology image of interest and extracts an embedding vector with meaningful information for classification. 
        $\mathcal{T}$ is to generate a class label as a text such as \textit{lymphocyte} and \textit{invasive carcinoma}. It obtains two inputs: visual input and text input. The visual input is an embedding vector from a pathology image, while the text input is a task-specific prompt that instructs the decoder to generate the relevant and proper prediction.
        $\mathcal{S}$ stores a set of adaptors that learn the task-specific representation in a resource- and computation-efficient manner. The overall architecture of the patch-level and slide-level CAMP is illustrated in Fig. \ref{fig_patch_workflow} and Fig. \ref{fig_slide_workflow}, respectively.
        
        For efficient and effective image classification, CAMP utilizes two types of knowledge: common knowledge and task-specific knowledge. The common knowledge is suitable for various tasks and is shared across different tasks. By contrast, the task-specific counterpart is utilized for a particular task, which is used in addition to the common knowledge to achieve a specialized capability for each classification task. In CAMP, the common knowledge is stored in $\mathcal{V}$ and $\mathcal{T}$, whereas task-specific knowledge is managed by  $\mathcal{S}$. The common knowledge is preserved by freezing corresponding modules, while the adaptors for the task-specific knowledge are trainable.

        \textbf{Visual encoder.}
            The role of the visual encoder $\mathcal{V}$ is to extract informative features in the form of an embedding vector given a pathology image. Any arbitrary CNN or Transformer-based models can be adopted and used as $\mathcal{V}$. Among various models, we consider three Transformer-based models, including CTransPath \cite{ctranspath}, Phikon \cite{phikon}, and UNI \cite{uni}, that are trained on a large number of pathology images in a self-supervised manner and shown to be effective in analyzing pathology images. 
            \textbf{CTransPath} is based on a 28M parameter SwinTransformer-Tiny \cite{swin1} with a patch partition layer replaced by a CNN. It was trained via a MoCoV3 \cite{mocov3} contrastive learning framework with diverse positive pairs sampled from different histopathology patches. The pretraining data includes about 15 million image patches from 32 thousand WSIs curated from TCGA (www.cancer.gov/tcga) and PAIP (http://www.wisepaip.org/paip).
            \textbf{Phikon} is an 86M parameter ViT-Base \cite{vit} that is pre-trained on approximately 6 thousand TCGA WSIs. The pretraining procedure is based on the iBOT \cite{ibot} contrastive learning framework with 43 million extracted patches. 
            \textbf{UNI} is built on a 307M parameter ViT-Large \cite{vit} on the in-house dataset Mass-100K with approximately 100 thousand WSIs. $\sim$ 100 million tiles are extracted for pretraining with the DINOv2 \cite{dinov2} contrastive objective.
            
        \textbf{Text decoder.}
            The text decoder $\mathcal{T}$ is responsible for pathology image classification in a generative fashion, given an image input and text input. The image input is an embedding $e$ processed by $\mathcal{V}$, which is adjusted by $\mathcal{S}$. The text input is a text prompt, such as ``the cancer grade of this prostate tissue is", used to guide $\mathcal{T}$ to generate a suitable prediction. This text prompt is converted by a tokenizer into tokens with the same dimension as the visual embedding $e$.
            These two inputs are then concatenated to form a final sequence.
            Given this sequence, $\mathcal{T}$ generates a class label in the form of a natural language term, such as \textit{well differentiated} or \textit{poorly differentiated}. The generation process is auto-regressive, i.e., $\mathcal{T}$ sequentially produces a new token based on previous tokens. 
            Similar to the visual encoder, we also employ the text decoder pretrained on pathology datasets to take advantage of rich in-domain knowledge. Although the generated text prediction in CAMP is shorter than other language tasks, such as image captioning or visual-question answering, the prediction contains specialized pathological words, e.g. carcinoma or lymphocyte, that are not exposed to general-domain language models. We employ 86M parameter \textbf{PLIP} \cite{plip} as the textual decoder $\mathcal{T}$, containing a stack of 12 Transformer encoder layers. PLIP was trained using OpenPath, a large-scale collection of approximately 200 million pathology image-text pairs curated from medical Twitter and other public sources.

        \textbf{Adaptor storage.}
            Both $\mathcal{V}$ and $\mathcal{T}$ are equipped with common knowledge in pathology acquired from a large collection of pathology data. Though such common knowledge can be utilized for various downstream tasks, one still needs to adapt to each task to further improve the performance. In other words, one needs to learn task-specific knowledge per downstream task. 
            Since there exist numerous downstream tasks, the adaptation process to each task should not interfere with other tasks, and task-specific knowledge should not revise the common knowledge.
            To this end, we design a dedicated component called adaptor storage $\mathcal{S}$ that allows us to learn task-specific knowledge. We construct $\mathcal{S}$ as a dictionary with task-specific $key$-$value$ pairs. Each classification task has a unique $key$ $\mathcal{K}$, represented as a trainable embedding vector. Each $\mathcal{K}$ is associated with a $value$ comprising a set of adaptors to tune the classification model to each downstream task. 
            These adaptors facilitate easy adaptation to a particular downstream task with the corresponding task-specific knowledge while preserving the common knowledge of $\mathcal{V}$ and $\mathcal{T}$. Hence, this decouples the optimization procedure of $\mathcal{V}$ and $\mathcal{T}$ from the adaptation procedure per task, and thus it prevents catastrophic forgetting (overwriting common knowledge with task-specific knowledge), allowing CAMP to effectively learn and conduct a variety of classification tasks. The composition of the adaptors differs between the patch-level and slide-level classifications.
    
            For patch-level classification, the adaptor set includes a visual encoder adaptor $\mathcal{S}_E$, a text decoder adaptor $\mathcal{S}_D$, and a projector adaptor $\mathcal{S}_P$. $\mathcal{S}_E$ and $\mathcal{S}_D$ are added to the original weights of $\mathcal{V}$ and $\mathcal{T}$ via low-rank adaptation (LoRA) \cite{hu2022lora}. 
            $\mathcal{S}_P$ serves as a connector that matches the embedding space of $\mathcal{V}$ with that of $\mathcal{T}$, ensuring the seamless alignment between $\mathcal{V}$ and $\mathcal{T}$. We build $\mathcal{S}_P$ using an efficient multiple-layered perceptron with four fully-connected layers. 

            For slide-level classification, the adaptor set comprises an aggregator adaptor $\mathcal{S}_A$, a text decoder adaptor $\mathcal{S}_D$, and a projector adaptor $\mathcal{S}_P$. For visual embedding, we utilize $\mathcal{V}$ only, which is fixed during the adaptation procedure following recent multiple instance learning (MIL) frameworks \cite{mil,mil2}. $\mathcal{S}_A$ is a parametric aggregator to combine patch embeddings into a single slide embedding in a trainable manner. $\mathcal{S}_D$ and $\mathcal{S}_P$ are the same as in the patch-level classification.
            
            To retrieve a suitable adaptor set for a given classification task, we devise a straightforward optimization and query generation mechanism. For each pair of an input image and text prompt, we generate a query by concatenating a visual embedding (from the visual encoder) and a textual embedding (from the text decoder).
            During training, the queries are employed to optimize $\mathcal{K}$ under two constraints. First, $\mathcal{K}$ should be similar to the query of the same classification task. Second, $\mathcal{K}$ should be far away from $\mathcal{K}$ of other tasks. The optimization of $\mathcal{K}$ is accomplished with a designated loss function $\mathcal{L_{\mathcal{K}}}$, which is described in Section 2.2. 
            At inference, the query is compared with all keys in the adaptor storage to retrieve the most suitable value, i.e. the most suitable adaptors for the classification task of interest.

        \textbf{Aggregator.}
             WSI classification is often formulated as a multiple-instance learning (MIL) problem \cite{mil}, a weakly supervised learning problem in which an aggregator is used to obtain a slide embedding from several patch embeddings. Following this, we employ the adapting aggregator $\mathcal{S}_A$ to generate a slide-level representation for WSI classification. We adopt four parametric aggregators from state-of-the-art MIL frameworks, including AB-MIL \cite{abmil}, CLAM-MB \cite{clam}, TransMIL \cite{shao2021transmil}, and IBMIL \cite{ibmil}. AB-MIL uses basic linear layers to predict attention scores for patch embeddings, uses these attention scores to compute the weighted combination of the embeddings, and generates a slide-level representation. CLAM-MB employs a multiple-branch attention aggregator where each branch is responsible for a classification class. It also learns an auxiliary classifier to identify distinguishable features between strongly and weakly attended patches. IBMIL utilizes a structured causal aggregator that conducts predictions at the bag level, mitigating confounders between bags and labels, and aims to uncover causal relationships and neutralize their influence through backdoor adjustments. TransMIL adopts a Transformer-based aggregator with a dedicated position encoding component called PPEG, which enables it to capture both morphological and spatial information of WSIs.

        \textbf{Low-rank adaptation.}
            We adopt LoRA \cite{hu2022lora} to adjust the weights of CAMP so as to conduct task-specific classification in an efficient and effective manner. Traditional finetuning methods adjust the entire weight matrix as follows $W_{new} = W + \delta W$ where $W \in \mathbb{R}^{d \times k}$ is the weight matrix and $\delta W \in \mathbb{R}^{d \times k}$ represents the amount of adjustment. Assuming that models have a low intrinsic dimension, LoRA decomposes the (large) weight matrix into smaller matrices as follows $W_{new} = W + \delta W = W + A \times B$ where $A \in \mathbb{R}^{d \times r}$ and $B \in \mathbb{R}^{r \times k}$ are the low dimension weight matrices that approximate $\delta W$ and $r \ll min(d,k)$. 
            LoRA can be utilized for any weight matrices in a model; however, we only apply LoRA to the projection matrices of the self-attention mechanism in the Transformer layers. We adjust the three matrices $W^q$, $W^k$, and $W^v$ that are used to calculate the query, key, and value in the attention mechanism, respectively. We note that the query, key, and value differ from the one in adaptor storage. For the rest of the paper, the query, key, and value are referred to as a component in the adaptor storage.

        \begin{figure}[!ht]%
            \centering
            \includegraphics[width=1\textwidth]{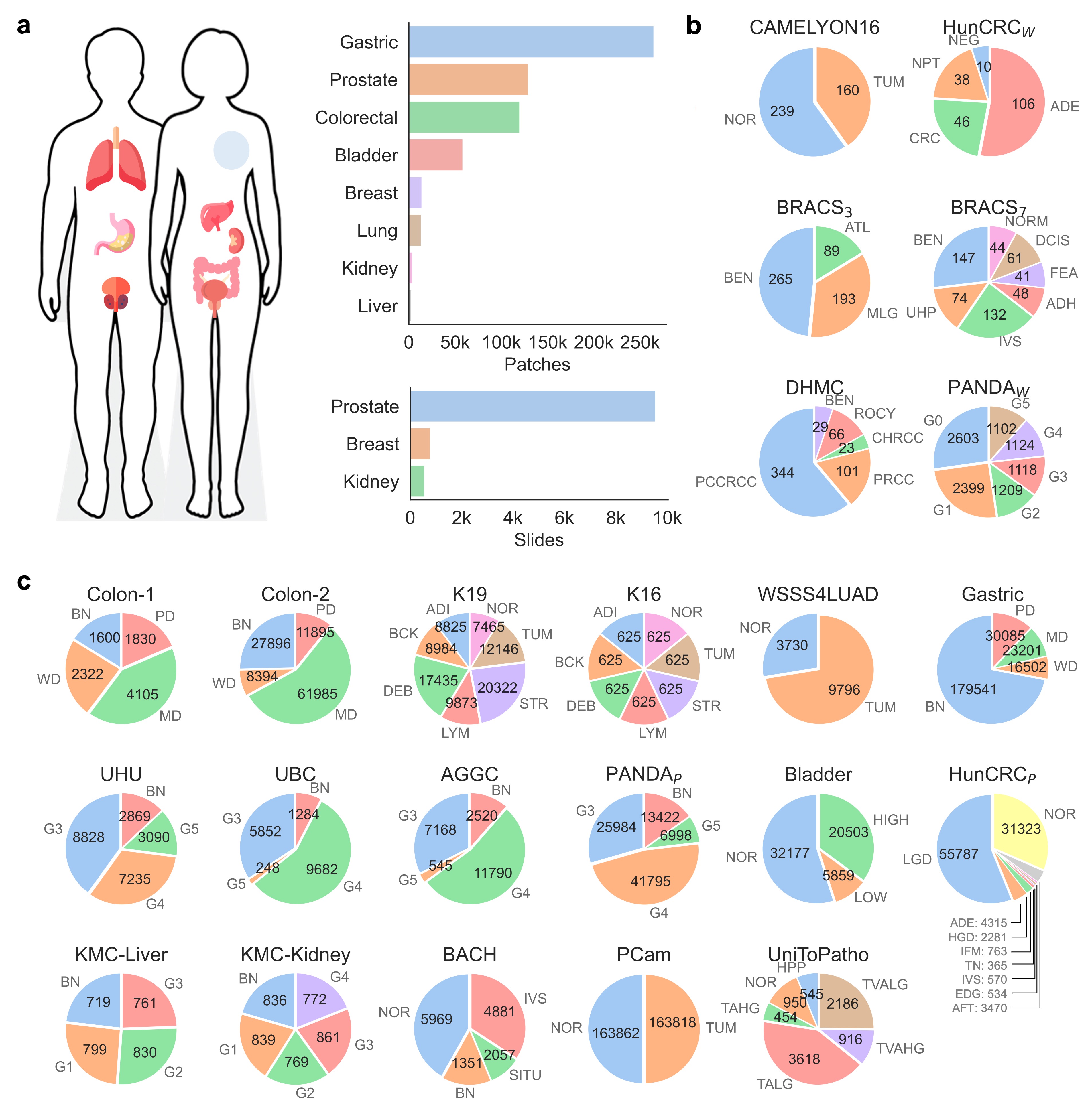}
            \caption{Datasets utilized for experiments. $\bold{a}$) 1,171,526 patches and 11,811 slides from 8 organs are curated for comprehensive experiments. $\bold{b}$) Class distribution of 6 slide-level datasets from 3 organs. $\bold{c}$) Class distribution of 17 patch-level datasets from 8 organs.}\label{fig_dataset}
        \end{figure}

        \begin{figure}[!ht]%
            \centering
            \includegraphics[width=1\textwidth]{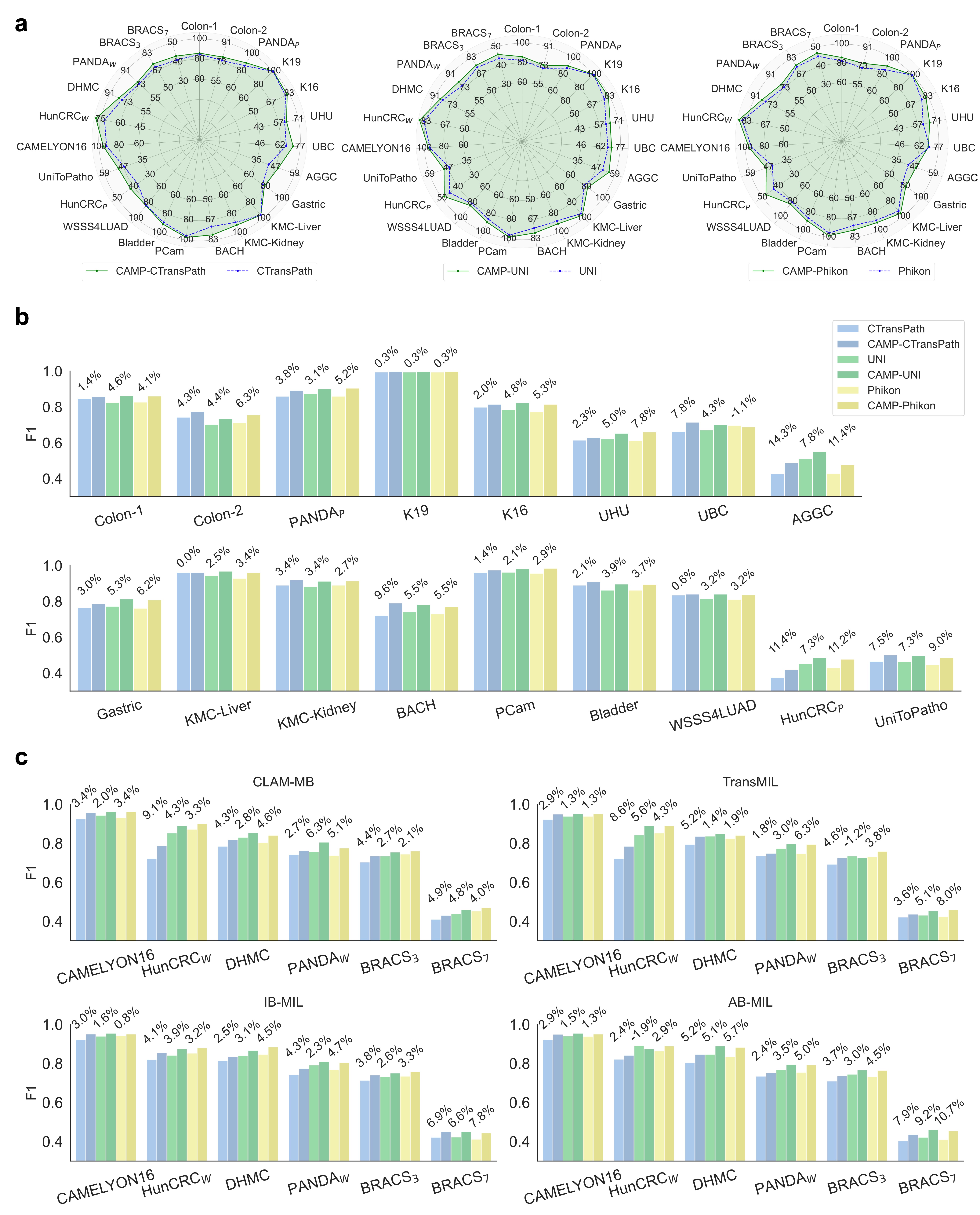}
            \caption{Performance of foundation models when integrated into CAMP. $\bold{a}$) CAMP increases the performance of CTransPath \cite{ctranspath}, UNI \cite{uni}, and Phikon \cite{phikon} on a wide range of datasets, on both patch- and slide-level classification. $\bold{b}$-$\bold{c}$) The detailed comparison in patch and slide datasets, respectively. The percentages show the ratios of change in the F1 score.}\label{fig_camp_vs_foundation}
        \end{figure}

        \begin{figure}[!ht]%
            \centering
            \includegraphics[width=1\textwidth]{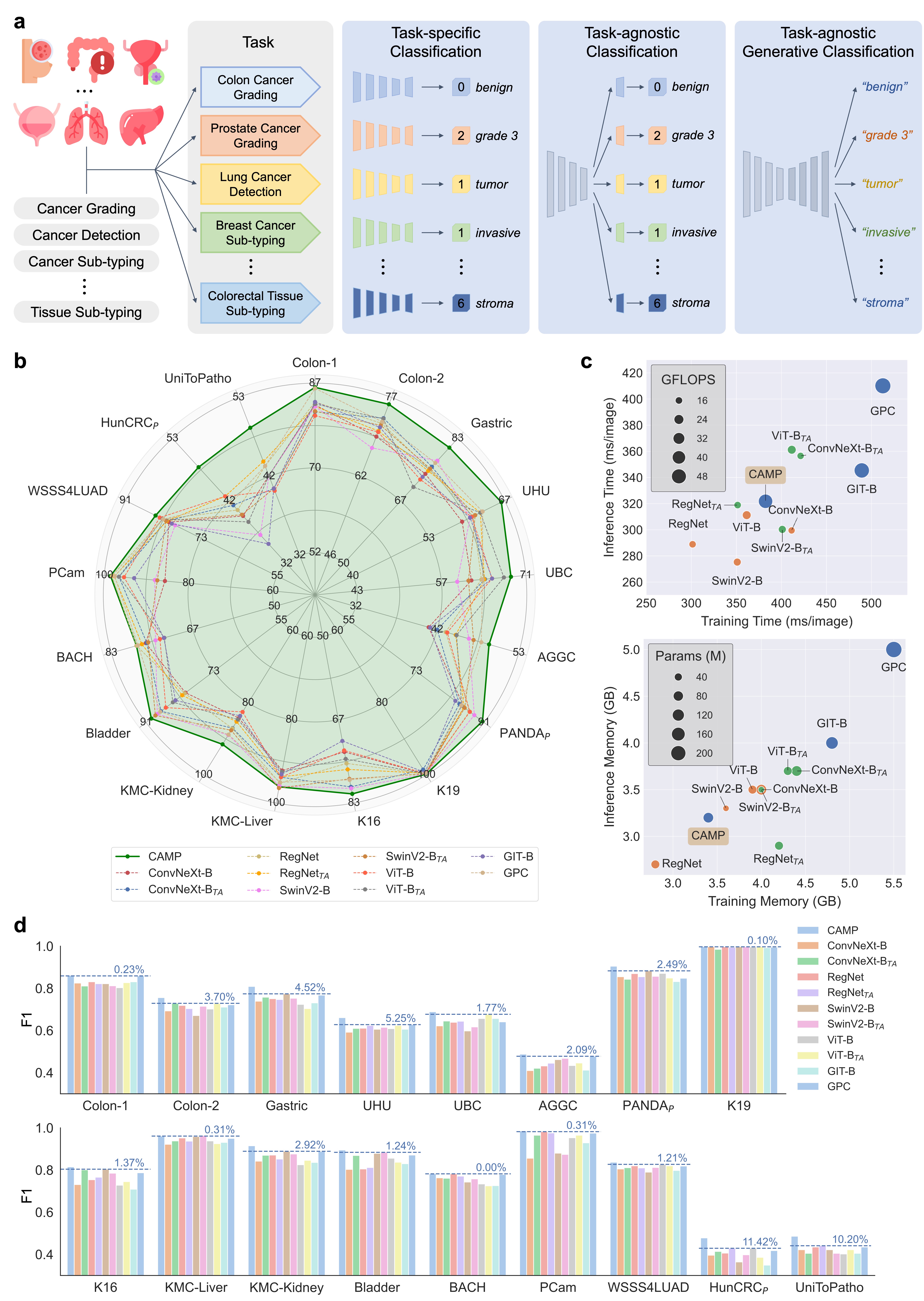}
            \caption{Comparison between the performance of efficiently finetuned CAMP and fully finetuned models. $\bold{a}$) The three configurations under consideration are task-specific, task-agnostic, and task-agnostic generative classification. $\bold{b}, \bold{d}$) CAMP performs better than other considered methods in 16/17 patch-level datasets. The numbers in the bar show the gap between CAMP and the second-best competitors. $\bold{c}$) Comparison between CAMP and competitors in terms of the computation time and memory consumption during training and inference.}\label{fig_camp_vs_full_ft}
        \end{figure}

        \begin{figure}[!ht]%
            \centering
            \includegraphics[width=1\textwidth]{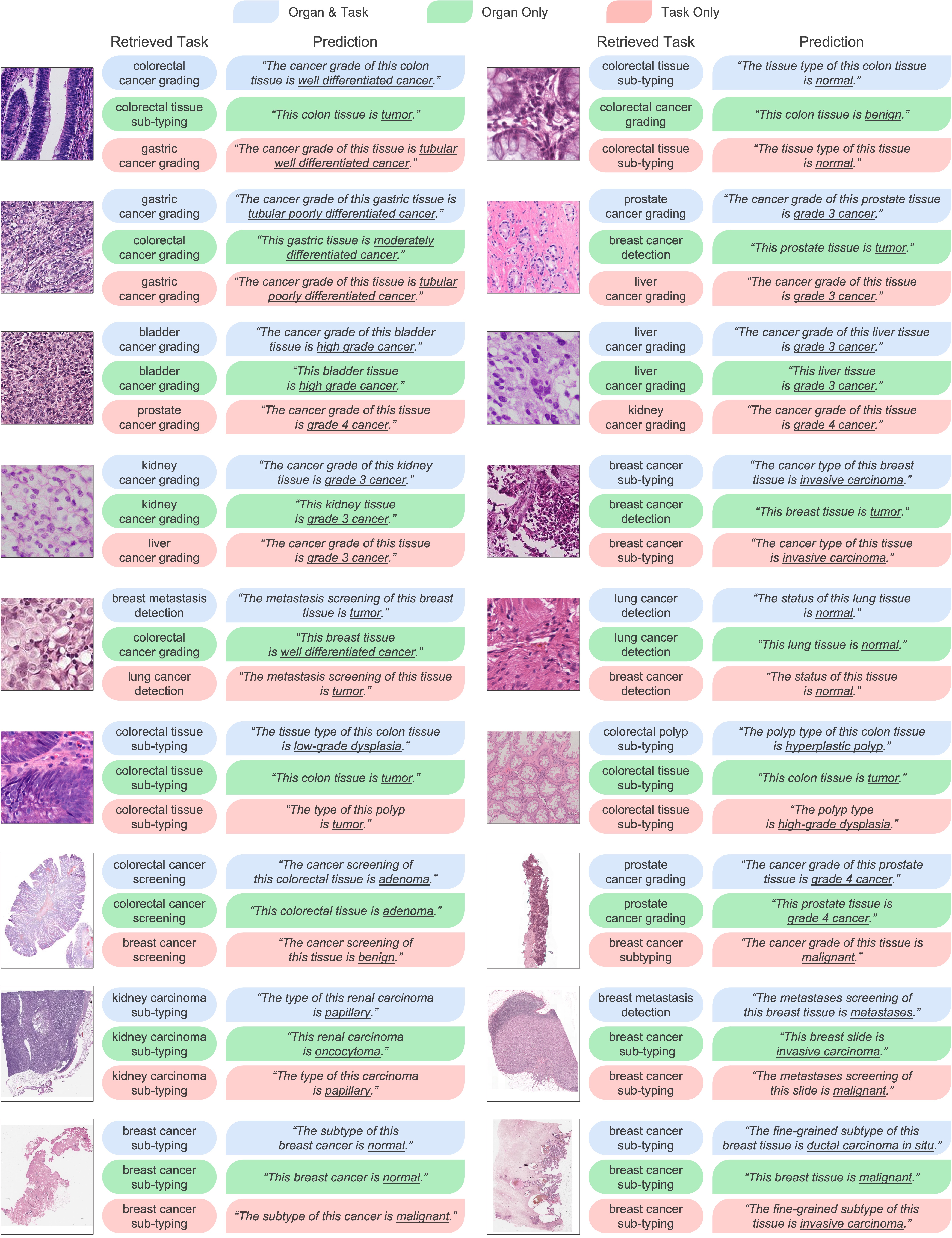}
            \caption{CAMP makes reasonable predictions regardless of the missing information in the text prompts.}\label{fig_ablation_missing_prompt}
        \end{figure}

        \begin{figure}[!ht]%
            \centering
            \includegraphics[width=1\textwidth]{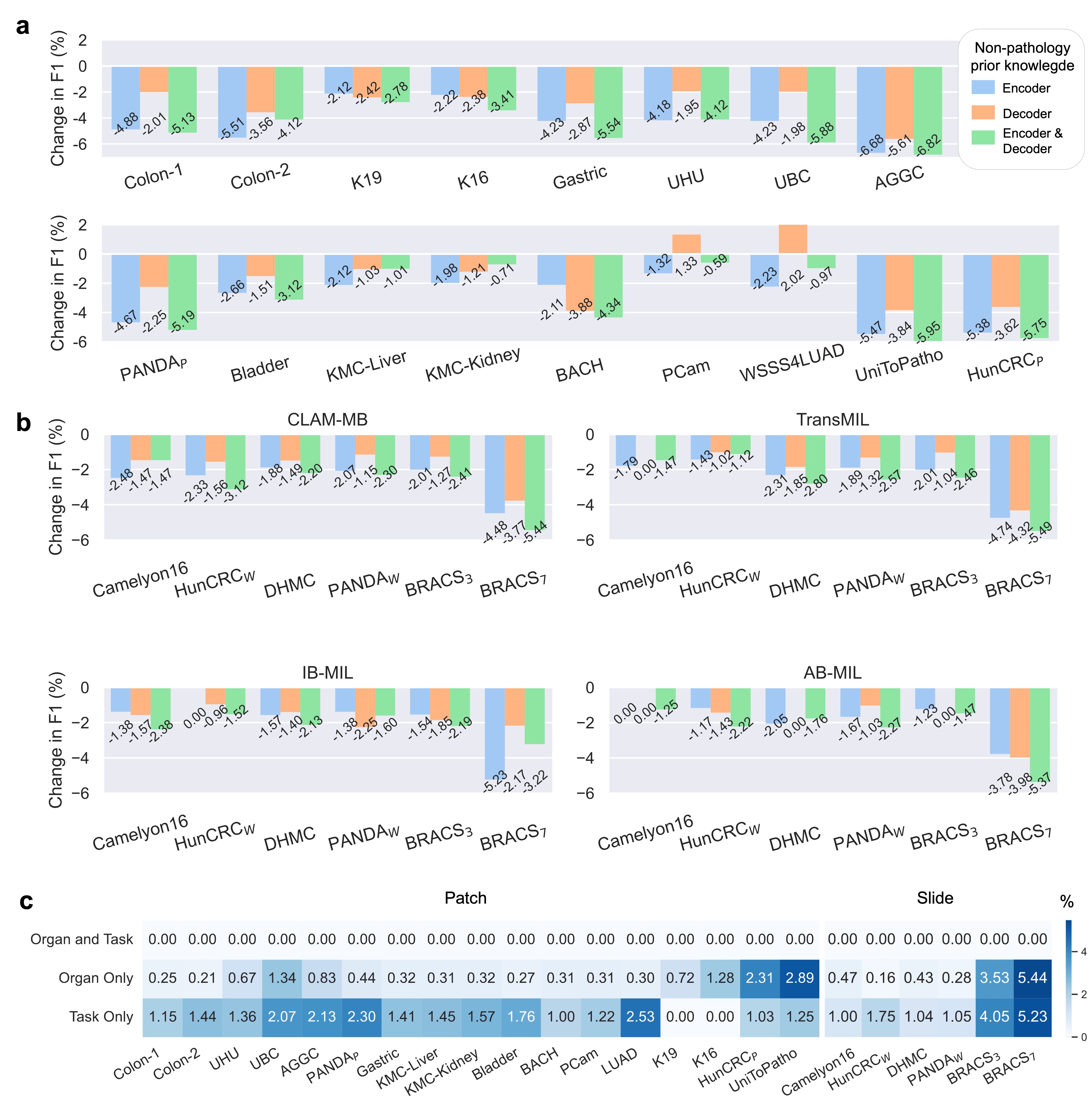}
            \caption{The importance of prior knowledge and text prompts. $\bold{a},\bold{b}$) The changes in F1 score when replacing pathology-pretrained modules with general-pretrained modules. These replacements only change the weights of this module while keeping the architectures. $\bold{c}$) The mis-retrieval rates when text prompts are missing information on the classification task.}\label{fig_ablation_pretraining_missing_rate}
        \end{figure}

        \begin{figure}[!ht]%
            \centering
            \includegraphics[width=1\textwidth]{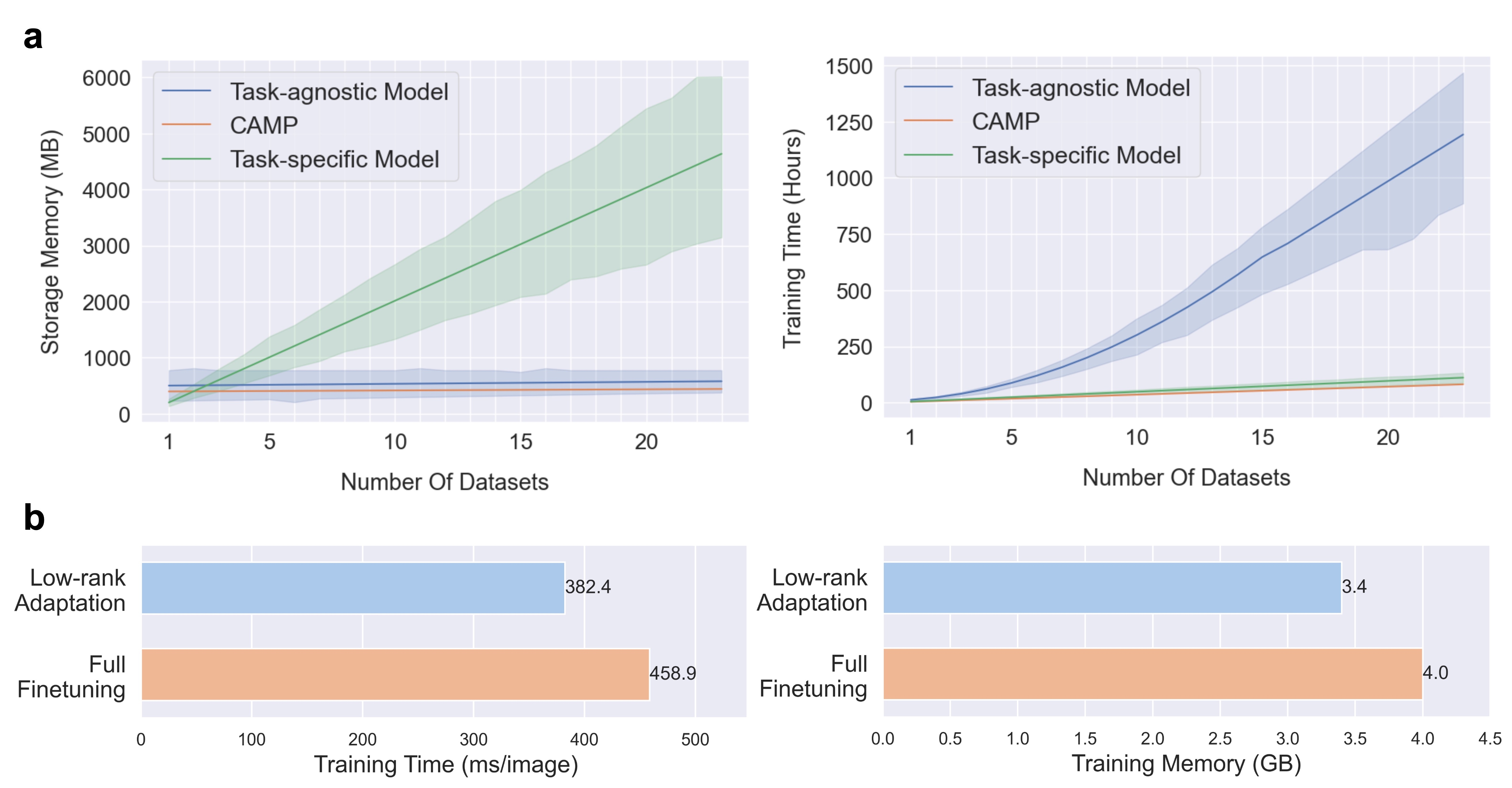}
            \caption{The efficiency of CAMP. $\bold{a}$) Scalability of CAMP with respect to storage memory and computation time as the number of datasets increases. $\bold{b}$) Training memory and time of low-rank adaptation and full finetuning.}\label{fig_lora_ft}
        \end{figure}

        \begin{figure}[!ht]%
            \centering
            \includegraphics[width=1\textwidth]{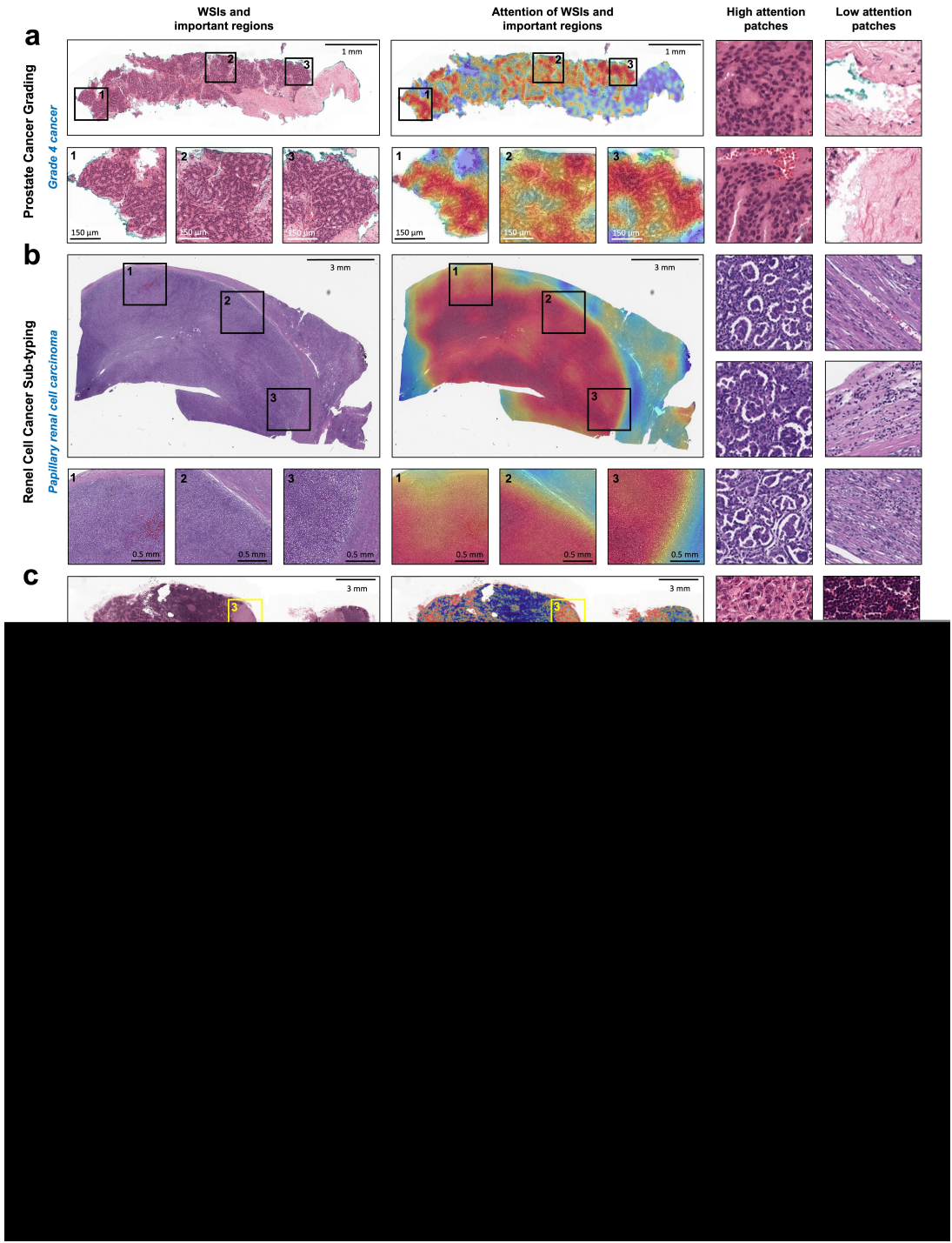}
            \caption{Interpretable heatmaps of CAMP. $\bold{a}$-$\bold{d}$) For each row, the name of the task and the prediction of CAMP are shown. The {\color{blue} blue texts} are similar to the ground truth, whereas the {\color{orange} orange text}  is different from the label. In the first two columns, a WSI and a generated attention heatmap are shown. The heatmaps represent the contribution of tissue patches to the final prediction. We show three important regions at the bottom of each WSI/heatmap, which are directly related to the diagnosis of the slide, e.g. tumors in breast cancer sub-typing. Additionally, patches with high and low scores are depicted on the last two columns for detailed observation.}\label{fig_heatmap}
        \end{figure}

        \begin{algorithm}
            \SetKw{Initialization}{Initialization} 
            \caption{CAMP training process}\label{training_algo}
            \KwIn{an image-label input $(x,y)$, a text prompt $z$, a pre-trained visual encoder $\mathcal{V}$, a pre-trained text decoder $\mathcal{T}$, keys of previous $M$-1 tasks $\{\mathcal{K}^{prev}_{i}\}_{i=1}^{M-1}$, and an adapting function $F$}

            \kwInit{A current key $\mathcal{K}^{cur}$ and adaptors (patch: $\mathcal{S}_E$, $\mathcal{S}_D$, $\mathcal{S}_P$; slide: $\mathcal{S}_A$, $\mathcal{S}_D$, $\mathcal{S}_P$).}

            $\mathcal{V} = freeze(\mathcal{V}) $                   \Comment{freeze visual encoder}

            $\mathcal{T} = freeze(\mathcal{T}) $                   \Comment{freeze text decoder}

            \eIf{type(x) = slide}
            {
                $ \{{x}_{i}\}_{i=1}^{N} = PatchExtract(x)$              \Comment{generate patch image bag}

                $ \{{e}_{i}\}_{i=1}^{N} = \bigl\{\mathcal{V}({x}_{i})\bigl\}_{i=1}^{N}$              \Comment{generate patch embedding bag}
    
                $ e_{v} = MaxPool \bigl(\{{e}_{i}\}_{i=1}^{N} \bigl)$ \Comment{generate slide visual embedding}
            }
            {
                $ e_{v} = \mathcal{V}(x) $ \Comment{generate patch visual embedding}
            }

            $ e_{t} = \mathcal{T}(z) $ \Comment{generate text embedding}
            
            $Q = Concat(e_{v}, e_{t})$            \Comment{generate query}

            \For{$epochs$}{
                \eIf{type(x) = slide}
                {
                    $ e_{v} = \mathcal{S}_A(\{{e}_{i}\}_{i=1}^{N})$   \Comment{extract slide visual embedding}
                }
                {
                    $\mathcal{V}'=F(\mathcal{V}, \mathcal{S}_E)$            \Comment{adapt visual encoder}
                    
                    $e_v = \mathcal{V}'(x)$                                \Comment{generate prediction}
                }
            
                $\mathcal{T}'=F(\mathcal{T}, \mathcal{S}_D)$            \Comment{adapt text decoder}
                
                $e_p = \mathcal{S}_P(e_v)$               \Comment{project visual embedding}
               
                $seq = Concat(e_p, z)$                     \Comment{generate input sequence}

                $\hat{y} = None $   \Comment{initialize prediction}
                
                \While{$\hat{y} \neq EOS$}{
                    $y' = \mathcal{T}'(seq)$           \Comment{generate prediction}
                
                    $seq = Concat(seq, Embedding(\hat{y}))$  \Comment{produce input embedding}
                
                    $\hat{y} = Append(\hat{y}, y') $    \Comment{update text output}
                }
    
                $\mathcal{L}_{\mathcal{K}} = - Sim(\mathcal{K}^{cur}, Q) + \sum_{i=1}^{M-1}Sim(\mathcal{K}^{cur}, \mathcal{K}^{prev}_i)$  \Comment{measure key loss}

                $\mathcal{L}_{\mathcal{S}} = CrossEntropy(y, \hat{y})$  \Comment{measure prediction loss}

                $\mathcal{L} = \mathcal{L}_{\mathcal{K}} + \mathcal{L}_{\mathcal{S}}$  \Comment{measure total loss}

                $\mathcal{L}.backprop()$        \Comment{update key and adaptors}
            }
            \KwOut{Optimal $\mathcal{K}^{cur}$ and adaptors.} 
        \end{algorithm}
    
    \subsection{CAMP training}        
        During training, we optimize the weights of $key$-$value$ pairs in the adaptor storage $\mathcal{S}$ by employing two loss functions $\mathcal{L}_{\mathcal{K}}$ and $\mathcal{L}_{\mathcal{S}}$ where $\mathcal{L}_{\mathcal{K}}$ is the loss for the key optimization and $\mathcal{L}_{\mathcal{S}}$ is the loss for the optimization of the adaptors ($\mathcal{S}_{E}$, $\mathcal{S}_{A}$, $\mathcal{S}_{P}$, and $\mathcal{S}_{D}$).
        The detailed illustration is shown in Fig. \ref{fig_patch_workflow}b, Fig. \ref{fig_slide_workflow}b, and Algorithm \ref{training_algo}. $\mathcal{L}_{\mathcal{K}}$ tries to pull the key $\mathcal{K}$ of the current task closer to the queries of the image-text prompt inputs (to learn the characteristic of the current task), while pushing it away from that of previous tasks (to clearly distinguish among classification tasks). The former is calculated by the dissimilarity of $\mathcal{K}$ and the queries, whereas the latter is the sum of similarity between $\mathcal{K}$ and previous keys. In this manner, $\mathcal{K}$ captures the task-related embedding in both visual and textual dimensions. We formally define $\mathcal{L}_{\mathcal{K}}$ as follow: 
        \begin{equation}
            \mathcal{L}_{\mathcal{K}} = - \dfrac {\mathcal{K}^{cur} \cdot Q} {\left\| \mathcal{K}^{cur}\right\|\left\| Q\right\|} + \dfrac{1}{M-1}\sum_{i=1}^{M-1} \dfrac {\mathcal{K}^{cur} \cdot \mathcal{K}^{prev}_i} {\left\| \mathcal{K}^{cur}\right\|\left\| \mathcal{K}^{prev}_i\right\|}
        \end{equation}\label{key_loss}
        where $Q$ is the query, $M$ is the number of tasks ($M-1$ tasks have been already examined), $\left\| \right\|$ denotes the Euclidean norm, $\mathcal{K}^{cur}$ is the key of the current task, and $\{\mathcal{K}^{prev}_i\}_{i=1}^{M-1}$ are the keys of the previous tasks.
        
        $\mathcal{L}_{\mathcal{S}}$ quantifies the correctness of the text output in comparison to the ground truth text label. It it used to update $\mathcal{S}$ only, while preserving the pre-trained weights of $\mathcal{V}$ and $\mathcal{T}$. 
        Given the token sequence generated by CAMP and the ground truth token sequence, $\mathcal{L}_{\mathcal{S}}$ aims to minimizing the difference between the probability distributions of the two token sequences. $\mathcal{L}_{\mathcal{S}}$ is formulated as follow:
        \begin{equation}
            \mathcal{L}_{\mathcal{S}} = -\sum_{i=1}^{N} y_{i}\log(p_{i})
        \end{equation}\label{value_loss}
        where $N$ is the size of the token sequence and $y_{i}$ and $p_{i}$ is the ground truth and output probability for the $i$th token, respectively.

        \begin{algorithm}
            \SetKw{Initialization}{Initialization} 
            \caption{CAMP inference process}\label{inference_algo}
            \KwIn{an image $x$, a text prompt $z$, a pre-trained visual encoder $\mathcal{V}$, a pre-trained text decoder $\mathcal{T}$, and an adaptor storage $\mathcal{S}$ with $M$ $key$-$value$ pairs $\{\mathcal{K}_{i}, \mathcal{S}_{i}\}_{i=1}^{M}$.}

            \eIf{type(x) = slide}
            {
                $ \{{x}_{i}\}_{i=1}^{N} = PatchExtract(x)$              \Comment{generate patch image bag}

                $ \{{e}_{i}\}_{i=1}^{N} = \bigl\{\mathcal{V}({x}_{i})\bigl\}_{i=1}^{N}$              \Comment{generate patch embedding bag}
    
                $ e_{v} = MaxPool \bigl(\{{e}_{i}\}_{i=1}^{N} \bigl)$ \Comment{generate slide visual embedding}
            }
            {
                $ e_{v} = \mathcal{V}(x) $ \Comment{generate patch visual embedding}
            }

            $ e_{t} = \mathcal{T}(z) $ \Comment{generate text embedding}
            
            $Q = Concat(e_{v}, e_{t})$            \Comment{generate query}

            $\mathcal{K} = {\underset{\mathcal{K}_i}{\arg\max}} \hspace{0.1cm} Sim(Q, \mathcal{K}_i)$  \Comment{select the most suitable key}

            \eIf{type(x) = slide}{
                $ (\mathcal{S}_A, \mathcal{S}_P, \mathcal{S}_D) = \mathcal{S}[K]$   \Comment{retrieve slide adaptors}

                $e_v = \mathcal{S}_A(\{{e}_{i}\}_{i=1}^{N})$           \Comment{extract slide visual feature}
            }
            {
                $ (\mathcal{S}_E, \mathcal{S}_P, \mathcal{S}_D) = \mathcal{S}[K]$   \Comment{retrieve patch adaptors}
                            
                $\mathcal{V}'=F(\mathcal{V}, \mathcal{S}_E)$            \Comment{adapt patch visual encoder}
                
                $e_v = \mathcal{V}'(x)$           \Comment{extract visual feature}
            }

            $\mathcal{T}'=F(\mathcal{T}, \mathcal{S}_D)$            \Comment{adapt text decoder}
            
            $e_p = \mathcal{S}_{P}(e_v)$               \Comment{project visual feature}

            $seq = Concat(e_p, z)$                   \Comment{generate input sequence}

            $\hat{y} = None $   \Comment{initialize prediction}
            
            \While{$\hat{y} \neq EOS$}{
                $y' = \mathcal{T}'(seq)$           \Comment{generate prediction}
            
                $seq = Concat(seq, Embedding(\hat{y}))$  \Comment{produce input embedding}
            
                $\hat{y} = Append(\hat{y}, y') $    \Comment{update text output}
            }

        \KwOut{Text output $\hat{y}$.} 
        \end{algorithm}

    \subsection{CAMP inference}
        The inference of CAMP can be split into two phases, including the retrieval of task-specific adaptors and the generation of the text output. In the first phase, the image input and text prompt input are fed into $\mathcal{V}$ and $\mathcal{T}$, respectively, and the resultant embedding vectors are concatenated to generate a query. The query is compared against all the keys in the adaptor storage, and the most similar $key$-$adaptors$ pair is retrieved. In the second phase, the retrieved adaptors are integrated into the generative classification model to effectively adapt to the inference task. Then, the image input and text prompt input are forwarded to the adapted classification model to produce text tokens in an auto-regressive fashion. 
        Specifically, the input image goes through $\mathcal{V}+\mathcal{S}_E$ (patch-level)/$\mathcal{V}+\mathcal{S}_A$ (slide-level) and $\mathcal{S}_P$ to produce the image embedding vector. The text prompt input is processed by $\mathcal{T}$ to generate the text embedding vector. The two input embedding vectors are then concatenated, forming an input embedding vector, and fed into $\mathcal{T}+\mathcal{S}_D$ to generate a new text token. The embedding vector of the new text token is concatenated with the input embedding vector and is used to generate the next text token. This process is repeated until it generates the EOS (end-of-sequence) token. The inference process is demonstrated in Fig. \ref{fig_patch_workflow}c, Fig. \ref{fig_slide_workflow}c, and Algorithm \ref{inference_algo}.

\section{Experiments}\label{sec2}
    \subsection{Datasets}
        We employ 22 datasets from 8 organs, including colorectal, gastric, lung, breast, kidney, prostate, bladder, and liver tissues, for pathology image classification (Fig. \ref{fig_dataset}). There exist 17 classification tasks that are categorized into 5 categories such as cancer grading, metastasis detection, cancer sub-typing, tissue sub-typing, and polyp sub-typing.

        \textbf{Colorectal cancer grading:} Two public datasets (\textbf{Colon-1} and \textbf{Colon-2}) are collected from \cite{joint}. Colon-1 contains 9,857 patch images obtained from 3 WSIs and 6 tissue microarrays (TMAs), scanning at 40x magnification by an Aperio digital slide scanner (Leica Biosystems). 
        Colon-2 has 110,170 patch images derived from 45 WSIs, digitized at 40x magnification using a NanoZoomer digital slide scanner (Hamamatsu Photonics K.K).
        Colon-1 is split into training (7,027), validation (1,242), and test set (1,588). Colon-2 is utilized as an independent test set. Each patch image has a spatial size of 512 $\times$ 512 pixels and is assigned a class label, including \textit{benign}, \textit{well differentiated cancer}, \textit{moderately differentiated cancer}, and \textit{poorly differentiated cancer}. We use \textit{``The cancer grade of this colon tissue is''} as the text prompt for the colorectal cancer grading task.

        \textbf{Prostate cancer grading:} Five public datasets are utilized for prostate cancer grading. 
        The first set (\textbf{UHU}), acquired from the Harvard Dataverse (https://dataverse.harvard.edu/), includes 22,022 image patches of size 750 $\times$ 750 extracted from 5 TMAs with 886 tissue cores. These 5 TMAs were digitally scanned at 40x magnification using a NanoZoomer digital slide scanner (Hamamatsu Photonics K.K.) at the University Hospital Zurich. 
        The second dataset (\textbf{UBC}) is the training set of the Gleason2019 challenge (https://gleason2019.grand-challenge.org/). This dataset comprises 17,066 image patches of size 690 $\times$ 690 from 244 prostate tissue cores, and each core was digitally scanned at 40x magnification using an Aperio digital slide scanner (Leica Biosystems).  
        The third set (\textbf{AGGC} \cite{aggc}) includes 22,023 image patches of size 512 $\times$ 512 obtained from WSIs of prostatectomy and biopsy specimens scanned at 20x magnification using multiple scanners including Akoya Biosciences, Olympus, Zeiss, Leica, KFBio, and Philips. 
        The last two datasets are obtained from the PANDA challenge \cite{panda}. 
        The fourth dataset, \textbf{PANDA}$_{\textbf{W}}$ \cite{panda}, is the slide-level classification dataset which includes 10,616 WSIs digitized at 20x magnification using a 3DHistech Pannoramic Flash II 250 scanner. Among them, we utilize 9,555 high-quality WSIs following \cite{uni}. 
        The fifth dataset, \textbf{PANDA}$_{\textbf{P}}$, is the patch-level classification dataset derive from \textbf{PANDA}$_{\textbf{W}}$, including 88,199 patch image of size 512 $\times$ 512. 
        All the image patches and WSIs are labeled with four classes: \textit{benign}, \textit{grade 3 cancer}, \textit{grade 4 cancer}, and \textit{grade 5 cancer}. 
        UHU is divided into training (15,303), validation (2,482), and test set (4,237). \textbf{PANDA}$_{\textbf{P}}$ is split into training (53,479), validation (17,023), and test (17,697) sets. \textbf{PANDA}$_{\textbf{W}}$ is split into training (7,647), validation (954), and test (954) sets. UBC and AGCC are adopted as independent test sets for the patch-level classification. The text prompt for this task is \textit{``The cancer grade of this prostate tissue is''}. 

        \textbf{Gastric cancer grading:} We utilize a public dataset \textbf{Gastric} \cite{camel} comprising 98 WSIs of 98 patients, which was digitized at 40x magnification using an Aperio digital slide scanner (Leica Biosystems). A total of 265,066 image patches, each with a spatial size of 512 $\times$ 512 pixels, are extracted and annotated with four class labels, including \textit{benign}, \textit{tubular well-differentiated cancer}, \textit{tubular moderately-differentiated cancer}, and \textit{tubular poorly-differentiated cancer}. The entire dataset is partitioned into a training (233,898), a validation (15,381) , and a test set (15,787). The text prompt for this task is \textit{``The cancer grade of this gastric tissue is''}. 

        \textbf{Bladder cancer grading:} A public bladder dataset \textbf{Bladder} \cite{bladder}, comprising 913 WSIs that are scanned at 40x magnification, is employed for bladder cancer grading. This consists of 58,539 patch images of size 1024 $\times$ 1024 that are extracted and split into a training (26,450), validation (12,912), and testing set (19,177). The patch images are categorized into 3 classes: \textit{normal}, \textit{low-grade cancer}, \textit{high-grade cancer}. \textit{``The cancer grade of this bladder tissue is''} is used for the text prompt for bladder cancer grading. 

        \textbf{Liver cancer grading:} A public dataset for liver cancer grading is collected from \cite{kmcliver}, denoted as \textbf{KMC-Liver}. This comprises 3,109 patch images of size 214 $\times$ 214 pixels that were initially obtained from 257 WSIs. The images are categorized into four sub-types of liver Hepatocellular Carcinoma (HCC) tumors: \textit{benign}, \textit{grade 1 cancer}, and \textit{grade 2 cancer}, and \textit{grade 3 cancer}. The entire dataset is utilized for training (2,549), validation (280), and testing (280) with \textit{``The cancer grade of this liver tissue is''} as the text prompt.

        \textbf{Kidney cancer grading:} We collect a kidney cancer grading dataset (\textbf{KMC-Kidney}) from \cite{kmckidney}, comprising 4,077 patch images of size 224 $\times$ 224 pixels. The patch images were initially obtained from surgical biopsies of kidney tissues. Each image is classified into five categories: \textit{benign}, \textit{grade 1 cancer}, and \textit{grade 2 cancer}, \textit{grade 3 cancer}, and \textit{grade 4 cancer}. The entire dataset is divided into training (3,432), validation (503), and test set (142). The text prompt for kidney cancer grading is \textit{``The cancer grade of this kidney tissue is''}.

        \textbf{Colorectal tissue sub-typing:} We employ four publicly available datasets for colorectal tissue sub-typing. 
        The first dataset, \textbf{K19} \cite{k19}, comprises 100,000 20x-digitized images of size 244 $\times$ 224 pixels from 9 tissue classes, whereas the second dataset, \textbf{K16} \cite{k16}, consists of 5,000 images sized at 150 $\times$ 150 pixels with 8 classes. Following \cite{abbet2021selfrule}, we match the number of classes between K19 and K16 by excluding one class (625 \textit{complex stroma} images) from K16 and by grouping stroma/muscle and debris/mucus into stroma and debris, respectively, in K19, resulting in 7 classes for both. The 7 classes are \textit{adipose}, \textit{background}, \textit{debris}, \textit{lymphocyte}, \textit{normal}, \textit{stroma}, and \textit{tumor}. K19 is utilized for training (70,000), validation (15,000), and testing (15,000), whereas K16 is used as an independent test set. 
        Moreover, we utilize HunCRC \cite{huncrc} as the third (\textbf{HunCRC}$_{\textbf{W}}$) and fourth (\textbf{HunCRC}$_{\textbf{P}}$) datasets. 
        \textbf{HunCRC}$_{\textbf{W}}$ is the slide-level classification dataset with 200 WSIs scanned at 20x magnification that are annotated with 4 classes \textit{negative}, \textit{non-neoplastic lesion}, \textit{carcinoma}, and \textit{adenoma}. 
        \textbf{HunCRC}$_{\textbf{P}}$ is the patch-level classification dataset, including 101,398 patch images of size 512 $\times$ 512 pixels. The patch images are classified into 9 categories: \textit{adenocarcinoma}, \textit{high-grade dysplasia}, \textit{low-grade dysplasia}, \textit{inflammation}, \textit{tumor necrosis}, \textit{suspicious for invasion}, \textit{resection edge}, \textit{technical artifacts}, and \textit{normal}. Both datasets are divided into training, validation, and test sets such as 158, 21, and 21 WSIs for \textbf{HunCRC}$_{\textbf{W}}$ and 81,118, 10,140, and 10,140 patch images for \textbf{HunCRC}$_{\textbf{P}}$, respectively.
        The prompt for these datasets is \textit{``The tissue type of this colon tissue is''}. 

        \textbf{Colorectal polyp sub-typing:} We employ \textbf{UniToPatho} \cite{unitopatho} for the classification of colorectal polyps. The dataset includes 9,536 patch images of size 1,812 $\times$ 1,812 pixels, scanned at 20x magnification. The images are grouped into 6 sub-types: \textit{normal}, \textit{hyperplastic polyp}, \textit{tubular adenoma with high-grade dysplasia}, \textit{tubular adenoma with low-grade dysplasia}, \textit{tubulo-villous ade-noma with high-grade dysplasia}, and \textit{tubulo-villous ade-noma with low-grade dysplasia}.
        The training, validation, and test sets include 6,329, 560, and 2,647 patch images, respectively. The prompt for UniToPatho is \textit{``The polyp type of this colon tissue is''}.

        \textbf{Kidney cancer sub-typing:} We utilize \textbf{DHMC} \cite{dhmc} for the 5-class renal cell carcinoma classification, including \textit{oncocytoma}, \textit{chromophobe}, \textit{clear cell}, \textit{papillary}, and \textit{benign}. The dataset consists of 563 WSIs, originally scanned by an Aperio AT2 whole-slide scanner at 20x magnification, and is split into training (393), validation (23), and testing (147) sets. The prompt for DHMC is \textit{``The subtype of renal cell carcinoma is''}. 

        \textbf{Breast cancer sub-typing:} We employ two public datasets. The first dataset, \textbf{BACH}, is obtained from Grand Challenge on Breast Cancer Histology Images \cite{bach}. This dataset comprises 14,258 patch images of size 512 $\times$ 512 pixels digitized at 20x magnification. Each image is annotated with one of the following four classes:  \textit{normal tissue}, \textit{benign}, \textit{in situ carcinoma}, and \textit{invasive carcinoma}, which were unanimously determined by two pathologists. We split them into training (8,752), validation (2,674), and test (2,832) sets. We use \textit{``The cancer type of this breast tissue is''} as the text prompt for this task. We adopt the second dataset, \textbf{BRACS}, from \url{www.bracs.icar.cnr.it} for the slide-level breast carcinoma classification. The dataset includes 547 WSIs collected from 189 patients with two different ways of labeling. The coarse subtyping includes 3 classes: \textit{benign tumor}, \textit{atypical tumor}, and \textit{malignant tumor}, whereas the 7-way fine-grained categories are \textit{normal}, \textit{pathological benign}, \textit{usual ductal hyperplasia}, \textit{flat epithelial atypia}, \textit{usual ductal hyperplasia}, \textit{ductal carcinoma in situ}, and \textit{invasive carcinoma}. The dataset is divided into training (395), validation (65), and testing (87) sets. The text prompts are \textit{``The subtype of this breast cancer is''} for the coarse-grained task and \textit{``The fine-grained subtype of this breast cancer is''} for the fine-grained task. 

        \textbf{Breast metastasis detection:} We utilize two public datasets (one for slide-level and the other for patch-level) derived from the Camelyon16 Challenge \cite{c16}, which are labeled with \textit{normal} and \textit{tumor}. The slide dataset, denoted as \textbf{CAMELYON16}, comprises 400 WSIs, digitized at 40x magnification, of sentinel lymph node sections. These slides are split into training (243), validation (27), and test (129) sets, excluding one mislabeled slide.
        The patch dataset, called \textbf{PCam}, has 327,680 patch images of size 96 $\times$ 96 pixels. The entire images are split into training (262,144), validation (32,768), and test (32,768) sets. The text prompt used for both datasets is \textit{``The metastasis screening of this breast tissue is"}.

        \textbf{Lung cancer detection:} We use \textbf{WSSS4LUAD} \cite{wsss} for the lung cancer detection task. This dataset consists of 97 WSIs digitized at 10x magnification. Initially, the dataset includes pixel-level semantic segmentation masks for tumor epithelial tissue, tumor-associated stroma tissue, and normal tissue. Using these masks, 13,526 patch images of size 224 $\times$ 224 pixels are extracted. These images are divided into training (10,091), validation (1,372), and test set (2,063). \textit{"The status of this lung tissue is"} is the text prompt for this task.

    \subsection{Experimental settings}
        To systematically evaluate CAMP, we integrate three foundation models in computational pathology (Phikon \cite{phikon}, CTransPath \cite{ctranspath}, and UNI \cite{uni}) into CAMP to verify the effectiveness of the proposed framework in comparison to standalone vision classification models. 
        Each foundation model is utilized as the visual encoder $\mathcal{V}$ in CAMP. The text decoder $\mathcal{T}$ is obtained from PLIP \cite{plip}. We strictly follow the original works \cite{phikon,ctranspath, uni, plip} to utilize the pre-trained weights and to pre-process data.
        Among the three foundation models, we select the best version of CAMP (CAMP-Phikon) and compare it against 6 deep learning models to further investigate the effectiveness of CAMP. The 6 models can be categorized into three groups based on their architecture: 1) 4 deep vision models: ConvNeXt-B \cite{convnext}, RegNet \cite{regnet}, SwinV2-B \cite{swin2}, and ViT-B \cite{vit} 2) 2 generative models: GPC \cite{GPC} and GIT-B \cite{git}. All the models are pre-trained on general domain knowledge, e.g. ImageNet, for visual pre-training. All the pre-trained weights are obtained from PyTorch Vision (https://pytorch.org/vision/) and HuggingFace (https://huggingface.co/).
        
        We investigate CAMP and 6 deep learning models under three experimental settings (Fig. \ref{fig_camp_vs_full_ft}a). As a result, we compared CAMP with 10 classification models with 3 settings: 1) task-specific classification ($C_{TS}$): a model is constructed with a feature extractor and a classifier head. It is trained on a specific training set and then tested on the corresponding test set(s) for each classification task; 2) task-agnostic classification ($C_{TA}$): a model has a feature extractor and a number of classifier heads, of which each is dedicated to one classification task. The model is trained on the combined training sets from all the classification tasks and evaluated on each test set using the classifier head associated with the specific task; 3) task-agnostic generative classification ($C_{TAG}$): a model includes a feature extractor and a generative classifier. All CNN (ConvNeXt-B and RegNet) and Transformer (SwinV2-B and ViT-B) models are employed for $C_{TS}$ and $C_{TA}$. CAMP, GIT-B, and GPC are utilized for $C_{TAG}$. It is noticeable that GIT-B and GPC are trained on all datasets at once, while CAMP is optimized on each dataset separately. Hence, CAMP learns the task-specific knowledge in multiple training phases, while the other two models are fully fine-tuned on all tasks in a single training process. The computational complexity of CAMP and 10 competitors are available in Fig. \ref{fig_camp_vs_full_ft}c.

    \subsection{Training details}
        For patch-level classification tasks, we employ the original data processing of each model. The training epoch is set to 100 with an initial learning rate of 0.0001 and a batch size of 256. AdamW \cite{adamw} is utilized as an optimizer along with the cosine decay scheduler. For LoRA, two parameters $r$ and $alpha$ are set to 6 and 12, respectively. $dropout$ is used with a chance of 0.1. The dimension of hidden states in the projector is 1024, 4096, and 2048, with GeLU as an activation function. 

        As for slide-level classification tasks, we follow the original data processing of each model. The training epoch is 200 with early stopping. The learning rate is initially set to 0.0002 and is controlled by the cosine scheduler. Adam \cite{kingma2017adam} is used for the model optimization. The settings of the projector and LoRA parameters are the same as those in the patch-level classification tasks. 

    \subsection{Evaluation metrics}
        We employ various evaluation metrics depending on the properties of the class labels. For all the cancer grading and breast cancer sub-typing tasks, we adopt four evaluation metrics: Accuracy (\textit{Acc}), Accuracy of cancer classification ($Acc_c$): ratio of correctly classified cancer samples among all cancer samples, macro-averaged F1 ($F1$), and quadratic-weighted kappa ($K_w$). For the rest of the tasks, the following four evaluation metrics are utilized: \textit{Acc}, macro-averaged Precision ($Pre$), macro-averaged Recall ($Rec$), and $F1$.

\section{Results}
    \subsection{CAMP improves the performance of pathology foundation models on a wide range of patch- and slide-level classification tasks}
        To investigate the effectiveness of CAMP, three pathology foundation models, including CTransPath \cite{ctranspath}, Phikon \cite{phikon}, and UNI \cite{uni}, were employed and compared to the framework of CAMP on 22 datasets from 8 organs with 17 patch-level datasets (11 tasks with about 1.1 million images) and 5 slide-level datasets (6 tasks with nearly 12,000 WSIs). 
        In other words, each of the three foundation models was individually and independently fine-tuned per classification task via linear probing, while three CAMP models (CAMP-CTransPath, CAMP-Phikon, and CAMP-UNI) were built and optimized for the entire slide- and patch-level classification tasks by using the corresponding foundation model as $\mathcal{V}$.
        For CAMP models, the text decoder is adopted from PLIP \cite{plip}. The results were measured using F1, accuracy, quadratic-weighted kappa, precision, and recall. Here, we primarily evaluate the models using F1 since it can be shared among different types of tasks. The detailed results are shown in Supplementary Table 1-27.

        Fig. \ref{fig_camp_vs_foundation} demonstrates the performance of CAMP and three pathology foundation models on both slide- and patch-level datasets. Across all 17 patch-level datasets, it was noticeable that CAMP improves upon the performance of the pathology foundation models. In a head-to-head comparison, CAMP, on average, increased F1 by 4.41\% for CTransPath, 4.40\% for UNI, and 5.12\% for Phikon. 
        We observed that the effect of CAMP varied across the datasets. For example, for colorectal cancer grading, CAMP increased F1 by 1.4\%, 4.6\%, and 4.1\% for CTransPath, UNI, and Phikon, respectively, on Colon-1. F1 was further improved on Colon-2 such as +4.3\% for CTransPath, +4.4\% for UNI, and +6.3\% for Phikon. 
        As for colorectal tissue sub-typing (K19, K16, and HunCRC$_P$), the average improvement in F1 by CAMP was 0.3\%, 4.0\%, and 10.0\% for K19, K16, and HunCRC$_P$, respectively. 
        In regard to prostate cancer grading (UHU, UBC, AGGC, and PANDA$_P$), CAMP substantially enhanced the performance of the foundation models except for Phikon on UBC, where F1 was dropped by 1.1\% by CAMP-Phikon in comparison to Phikon; on AGGC, which is highly imbalanced toward grade-4 samples (more than 50\%), CAMP attained the greatest performance improvement in F1 by 14.3\%, 7.8\%, and 11.4\% for CTransPath, UNI, and Phikon, respectively.
        
        Moreover, across all 5 slide-level datasets, CAMP, in general, offered the superior performance gain for the three pathology foundation models regardless of the type of the aggregators. 
        Overall, using CAMP, the classification performance, measured by F1, was improved by 2.59\% for breast cancer detection (CAMELYON16), 4.15\% for colon tissue sub-typing (HunCRC-S), 3.69\% for kidney cancer sub-typing (DHMC), 3.85\% for prostate cancer grading (PANDA-S), 3.11\% for coarse-grained breast cancer sub-typing (BRACS-3), and 6.63\% for fine-grained breast cancer sub-typing (BRACS-7). There were only two exceptions where CAMP was inferior to the foundation model; F1 of CAMP-UNI decreased by 1.2\% and 1.9\% in comparison to that of UNI on BRACS$_3$ on HumCRC$_W$, respectively.
        Regarding the four aggregators, CAMP, on average, increased F1 by 4.12\%, 3.75\%, 3.83\%, and 4.31\% for CLAM-MB, TransMIL, IB-MIL, and AB-MIL, respectively. 
        
        The results on the patch- and slide-level classification tasks suggest that CAMP is capable of conducting a variety of classification tasks at both patch- and slide-levels with high accuracy, CAMP is able to improve upon the pathology foundation models across different datasets and tasks, CAMP is robust to the choice of $\mathcal{V}$ and/or aggregator, and thus CAMP can serve as a generic framework for classification tasks. 
        Among the three CAMP models (CAMP-CTransPath, CAMP-Phikon, and CAMP-UNI), the performance of CAMP-CTransPath was, in general, inferior to that of the other two models on both patch- and slide-level classification tasks. Comparing CAMP-Phikon and CAMP-UNI, the two models achieved comparable performance; however, the computational complexity and memory requirement were much more significant for CAMP-UNI since Phikon is based on 86M-param ViT-Base while UNI is built on 307M-param ViT-Large. Hence, we chose Phikon as the default visual encoder $\mathcal{V}$ for CAMP, i.e., CAMP-Phikon is used to further evaluate the effectiveness and efficiency of CAMP in comparison to other classification models under various settings.
    
    \subsection{CAMP outperforms fully fine-tuned vision models}
        We further evaluated the classification performance of CAMP on the 11 patch-level classification tasks (colorectal cancer grading, prostate cancer grading, gastric cancer grading, bladder cancer grading, liver cancer grading, kidney cancer grading, breast cancer sub-typing, colorectal tissue sub-typing, colorectal polyp sub-typing, breast metastasis detection, and lung cancer detection) from 8 organs. There are 13 datasets that were split into training, validation, and testing sets. Using them, we trained CAMP in a serial fashion. The trained CAMP is applied to 4 external datasets (1 for colorectal cancer grading, 2 for prostate cancer grading, and 1 for colorectal tissue sub-typing) to test the generalization ability of CAMP on unseen datasets. We compared CAMP with 10 classification models (4 deep vision models: ConvNeXt-B, RegNet, ViT-B, and SwinV2-B, 4 task-agnostic deep vision models: ConvNeXt-B$_{TA}$, RegNet$_{TA}$, ViT-B$_{TA}$, and SwinV2-B$_{TA}$, and 2 generative models: GPC and GIT-B). Fig. \ref{fig_camp_vs_full_ft}b and d show the comparison between CAMP and other competitors in terms of F1 on the 17 datasets. Detailed results of all evaluation metrics are reported in Supplementary Table 1-11. 

        Overall, CAMP was able to conduct the 11 different classification tasks in an accurate and consistent manner (Fig. \ref{fig_camp_vs_full_ft}), achieving 0.756$\sim$0.861 F1, 0.809 F1, 0.488$\sim$0.905 F1, 0.478$\sim$0.998 F1, 0.961 F1, 0.915 F1, 0.895 F1, 0.782 F1, 0.985 F1, 0.486 F1, and 0.838 F1 for colorectal cancer grading (Colon-1 and Colon-2), gastric cancer grading, prostate cancer grading (UHU, UBC, AGGC, and PANDA), colorectal tissue sub-typing (K19, K16, and HunCRC$_{P}$), liver cancer grading, kidney cancer grading, bladder cancer grading, breast cancer sub-typing, breast metastasis detection, colorectal polyp sub-typing and lung cancer detection, respectively. 
                
        CAMP outperformed the 4 task-specific competitors in 16 of 17 datasets; the exception is BACH (breast cancer sub-typing), where RegNet obtained an F1 of 0.782, whereas CAMP achieved an F1 of 0.771.  
        It was remarkable that CAMP is superior to the second-best task-specific models by 3.6\%$\sim$5.1\% in colorectal cancer grading, 4.5\% in gastric cancer grading, 2.5\%$\sim$8.2\% in prostate cancer grading, 0.1\%$\sim$11.4\% in colorectal tissue sub-typing, 0.2\% in liver cancer grading, 2.9\% in kidney cancer grading, 1.8\% in bladder cancer grading, 0.3\% in breast metastasis detection, 1.2\% in lung cancer detection, and 11.7\% in colorectal polyp sub-typing. We note that the second-best task-specific model varied depending on the datasets. This indicates that the performance of the task-specific models, which were fully fine-tuned for downstream tasks, are inconsistent across differing datasets and tasks, whereas CAMP permits reliable and superior performance on a wide range of tasks and datasets. 

        Furthermore, CAMP surpassed the 6 task-agnostic competitors across the 11 classification tasks except for liver cancer grading. We made similar observations; CAMP outperformed the second-best task-agnostic models by 0.2\%$\sim$4.7\% in colorectal cancer grading, 5.3\% in gastric cancer grading, 2.1\%$\sim$5.6\% in prostate cancer grading, 0.1\%$\sim$3.6\% in colorectal tissue sub-typing, 3.0\% in kidney cancer grading, 1.2\% in bladder cancer grading, 0.4\% in breast metastasis detection, 2.1\% in lung cancer detection, and 11.7\% in colorectal polyp sub-typing; the performance of the task-agnostic models was unsteady, and thus the second-best model differed from one dataset to another. It is worth noting that the task-agnostic models were trained on the entire collection of the training datasets from the 11 classification tasks. This implies that the vanilla framework of the task-agnostic models is sub-optimal and the superior performance by CAMP is not simply due to the usage of the large datasets.

    \subsection{Prior knowledge on pathology data plays a critical role}
        
        In CAMP, we employ the visual encoder $\mathcal{V}$ and the text decoder $\mathcal{T}$ that were pre-trained on a large pathology image data. $\mathcal{V}$ learned the pathology-specific knowledge from $\sim$43 million pathology images via contrastive learning \cite{phikon}, whereas $\mathcal{T}$ was trained on about 200,000 pathology images paired with text descriptions \cite{plip}. Therefore, CAMP was exposed to pathology data prior to the adaptation to downstream tasks, i.e., 11 classification tasks. To investigate the importance of the pathology-specific prior knowledge on CAMP, we conducted the classification tasks by replacing the weights of $\mathcal{V}$ and $\mathcal{T}$ with the weights obtained from the natural images and natural languages, which are designated as the general prior knowledge. Specifically, in the first experiment, the weights of $\mathcal{V}$ were substituted by those from ImageNet, producing $\mathcal{V}_{g}$, while the weights of $\mathcal{T}$ were kept the same. In the second experiment, we adopt the weights of the text decoder of CLIP \cite{clip}, which were pre-trained on 400 million natural image-text pairs, and used them as the weights for $\mathcal{T}$, assigned as $\mathcal{T}_{g}$, but retained $\mathcal{V}$. The last experiment employed $\mathcal{V}_{g}$ and $\mathcal{T}_{g}$, in which CAMP was only equipped with the general prior knowledge.

        In the absence of the pathology-specific prior knowledge, the classification performance in patch-level tasks generally dropped (Fig. \ref{fig_ablation_pretraining_missing_rate}a,b); for instance, the average performance drop for the patch-level classification tasks was -3.6\%, -2.6\%, and -3.8\% by employing $\mathcal{V}_{g}$, $\mathcal{T}_{g}$, and both $\mathcal{V}_{g}$ and $\mathcal{T}_{g}$, respectively. Similar observations were made for the slide-level classification tasks, in which F1 decreased by 2.1\% for $\mathcal{V}_{g}$, 1.5\% for $\mathcal{T}_{g}$, and 2.5\% for both $\mathcal{V}_{g}$ and $\mathcal{T}_{g}$. 
        On the examination of each dataset, we found that the adoption of both $\mathcal{V}_{g}$ and $\mathcal{T}_{g}$ consistently results in a reduction in the classification performance; however, the degree of reduction in the performance varied across the datasets; for example, in the patch-level tasks, the largest performance drop of -6.68\% was achieved in AGGC (prostate cancer grading) and, in PCam (breast metastasis detection), the least performance drop of -0.97\% was attained.
        We also observed that $\mathcal{V}$ plays a crucial role in the classification at both patch- and slide-levels. The performance drop by $\mathcal{V}_{g}$ was almost always larger than the drop by $\mathcal{T}_{g}$, especially for the patch-level classification tasks. This might be due to the way CAMP processes the inputs and predicts the class labels. The output of the visual encoder is directly used for the text generation, and thus the mis-interpretation of the input image by the visual encoder would provide incorrect information for the text generation by the text decoder. In other words, the better the visual encoder is, the better information the text decoder attains, leading to improved classification performance.
        
        Though the average performance was substantially dropped by $\mathcal{T}_{g}$, its effect was disproportionate across the classification tasks. For most of the tasks, CAMP with $\mathcal{T}_{g}$ resulted in the performance drop ranging from -1.03\% (KMC-Liver) to -5.61\% (AGGC) for the patch-level classification tasks and from -0.96\% (HunCRC$_W$ with IB-MIL) to -4.32\% (BRACS$_7$ with TransMIL) for the slide-level classification tasks. 
        For some cases, the adoption of $\mathcal{T}_{g}$ did not affect the slide-level tasks such as Camelyon16 by TransMIL and AB-MIL, DHMC by AB-MIL, and BRACS$_3$ by AB-MIL.
        It even increased the patch-level classification performance by 1.33\% and 2.02\% for breast cancer detection (PCam) and lung cancer detection (WSSS4LUAD), respectively. This is a contributory factor in the small decrease in the performance when CAMP employed both $\mathcal{V}_{g}$ and $\mathcal{T}_{g}$. The increase in the performance by $\mathcal{T}_{g}$ may be ascribable to the nature of the classification tasks and class labels. For PCam and WSSS4LUAD, there exist two labels only, including \textit{normal} and \textit{tumor}, of which each label is relatively short and simple. Other classification tasks usually have more class labels, the labels tend to be long and complicated, such as \textit{tubular poorly-differentiated cancer}, \textit{invasive carcinoma}, and \textit{lymphocyte}, and/or the labels are infrequently used in natural languages.

    \subsection{CAMP is robust to the variations in the text prompt}
        CAMP needs two inputs, including a pathology image and a text prompt. At inference, the two inputs serve two purposes: one is to retrieve the appropriate adaptors and the other is to generate the text output using the adaptors. For the accurate and reliable prediction, the accurate retrieval of the adaptors is a prerequisite. In order to assess the accuracy of the adaptor retrieval on the classification tasks, we conducted the following three experiments. We first computed the rate of mis-retrieval of the adaptors given the input image-text prompt pairs per task. Then, we repeated the same experiment in the absence of the task or organ information. For example, the breast cancer sub-typing task initially has the text prompt \textit{the cancer sub-type of this breast tissue is}, i.e., both organ and task information are available. In the following two experiments, the text prompt changed to \textit{this breast tissue is} and \textit{the cancer sub-type of this tissue is}. The former contains the organ information only, and the latter includes the task information only.

        Fig. \ref{fig_ablation_missing_prompt}c depicts the rate of mis-retrieval with varying text prompts. Provided with both organ and task information, CAMP retrieved the correct adaptors without failure for the entire classification tasks. Missing either the organ or task information resulted in minimal mis-retrieval rates regardless of the classification tasks. For the organ only, there was a mis-retrieval rate of 0.57\% on average, ranging from 0.22\% to 1.49\%. As for the task only, the mis-retrieval rate varied from 0.00\% to 2.79\% and averaged 1.43\% across the 17 classification tasks. These results indicate that CAMP is able to retrieve the correct adaptors even though it is provided with the incomplete text prompt, demonstrating the validity of the key optimization. 
        
        Furthermore, we investigated the effect of the incorrect retrieval of the adaptors by comparing the predicted text outputs in the three experiments. It is remarkable that CAMP was, in general, able to generate the correct or semantically related text outputs even though the adaptors from different tasks were employed (Fig. \ref{fig_ablation_missing_prompt}).
        For example, given a \textit{well differentiated cancer} pathology image for colorectal cancer grading, CAMP retrieved the adaptors from colorectal tissue sub-typing and gastric cancer grading for the text prompt with the organ only and the task only, respectively, and the corresponding text outputs were \textit{tumor} and \textit{tubular well differentiated cancer}, respectively. 
        Similarly, for the \textit{grade 3 cancer} pathology image in liver cancer grading, the adaptors from liver cancer grading (organ only) and kidney cancer grading (task only) were retrieved. Using these adaptors, CAMP generated \textit{grade 3 cancer} and \textit{grade 4 cancer} for the organ-only and task-only text prompts, respectively. 
        As for the \textit{normal} pathology images from colorectal tissue sub-typing and lung cancer detection, CAMP produced either \textit{normal} or \textit{benign} regardless of the text prompts.
        Overall, CAMP almost always predicted benign/normal pathology images as \textit{benign} or \textit{normal}. Tumor/cancer pathology images were classified as \textit{tumor} or similar type of cancer. Hence, CAMP is capable of addressing incomplete information and providing contextually relevant outputs. As CAMP is exposed to more diverse and related tasks (e.g., tissue sub-typing), the quality and relevance of the output would be improved, holding the potential to serve as a robust, unified pathology image classification model.

    \subsection{CAMP achieves efficiency in both computation and storage}
        There are numerous classification tasks in computational pathology. The more computational pathology tools we use in the clinics, the more computational resources we need to provide. AI in healthcare, in general, faces critical sustainability issues on computer power, energy, and storage with the increase in the size and complexity of models \cite{jia2023importance}. To understand and analyze the potential impact of CAMP and other competitors on the clinics, we examined the efficiency of CAMP and other competitors in terms of the model complexity and the computational and storage requirement, including the number of parameters, Giga floating-point operations per second (GFLOPS), training time and memory consumption, and inference time and memory consumption (Fig. \ref{fig_camp_vs_full_ft}c). The training and inference time were estimated in milliseconds per image.  
        
        Overall, the traditional classification models (CNN and Transformer models) usually required a less amount of parameters, GFLOPs, time, and memory for both training and inference in comparison to the generative classification models (CAMP, GPC, and GIT-B) (Fig. \ref{fig_camp_vs_full_ft}c). This is mainly because the generative classification models consist of two modules, one for encoding and the other for decoding. 
        Comparing the traditional classification models, Transformer models (MaxViT, SwinV2-B, ViT-B, PLIP-V, and CTransPath) were computationally more expensive than CNN models (ConvNeXt-B, EfficientNetV2-S, ResNet50, RegNet, and ResNeXt50). 
        Among the generative classification models, CAMP was shown to be the most efficient model with respect to the number of parameters, GFLOPS, and the time and memory consumption for training and inference. CAMP was also comparable to the recent CNN and Transformer models with respect to the training and inference time and memory consumption.
        
        However, the above measurements are valid as we consider a single task only, which ignores the practical and forthcoming issues in the digital pathology era. The more realistic scenario would involve a great deal of tasks that are entirely or partially conducted or aided by AI-driven tools. To analyze CAMP and other models from this perspective, we investigated the scalability of CAMP and others by measuring the training time and storage memory as the number of datasets (tasks) increases (Fig. \ref{fig_lora_ft}). Other competitors were grouped into two categories: one includes task-specific models, and the other contains task-agnostic models. 
        The more datasets or tasks we have, the more storage memory the task-specific models require. This is because a new model is needed every time a new dataset or task is given. However, the storage memory that the task-agnostic models and CAMP need was shown to be steady since these only use a single model with and without additional, tiny parameters. For the 22 datasets used in this study, CAMP could save up to 85\% of the storage memory as compared to the task-specific models. 
        As for the training time, the training time of the task-agnostic models was exponentially increasing, but that of the task-specific models and CAMP was slowly increasing. The exponential increase by the task-agnostic models is ascribable to the usage of all the datasets, not just the newly added dataset.
        CAMP was also able to reduce the training time up to 94\%. These observations suggest that CAMP is efficient in both computation time and storage memory, and other models (both task-agnostic and task-specific models) are inefficient in computation time or storage memory.

        In order to learn and adapt to a new task, CAMP introduces low-rank adaptation (LoRA), which keeps and freezes the original weight matrices and only learns the amount of the additive adjustments to the weight matrices. LoRA decomposes each of the adjusted weight matrices into two low-dimensional weight matrices with a lower rank and a smaller number of trainable parameters. The traditional methods often adopt finetuning approaches that directly adjust the original weight matrices, and thus a new set of weight matrices is needed for each task, leading to a substantial increase in the number of parameters.
        To investigate the efficiency and effectiveness of LoRA, we trained and tested CAMP on the 17 classification tasks (11 patch-level and 6 slide-level) using the two approaches (LoRA and full fine-tuning). Then, we compared the training time and storage memory between the two approaches. 
        We note that, in LoRA, we only adjusted a small portion of the weight matrices, i.e., the projection matrices for self-attention in the Transformer layers. As for the full finetuning, the entire weight matrices in CAMP were independently adjusted per classification task, and thus this can be considered task-specific. The amount of storage memory for the full finetuning continuously grows as the number of tasks/datasets increases, while LoRA needs a tiny amount of additional storage memory for a new task.
        On the examination of the training time and memory, the efficiency of LoRA was evident. On average, LoRA required 382.4 milliseconds and 3.4 GB of memory to process an image during training, which saves 16.7\% of training time and 15.0\% of training memory as compared to the full finetuning (Fig. \ref{fig_lora_ft}b). This leads to a significant reduction in power and memory consumption as well as processing time during the development of the classification models, thereby shorting the time for the deployment to the clinics.

    \subsection{CAMP attends to critical regions}
        To deepen our understanding of the diagnosis of CAMP, we visualized and interpreted the relative importance of differing regions in the pathology slides using the attention weights of the aggregator. The attention weights represent the relative contribution of the corresponding patches in generating the slide-level embedding. Using the attention weights, we generated the attention heatmaps by converting the attention weights into percentiles, normalizing them, and plotting the normalized scores as color-maps using the corresponding patch coordinates. Following \cite{clam}, we generated fine-grained attention heatmaps by overlapping the regions/patches and averaging the attention scores.
        The exemplary WSIs and the corresponding heatmaps are depicted in Fig. \ref{fig_heatmap}. For each pair of a WSI and a heatmap, we show three highly attended regions and 2-3 patches that receive high and low attention at high magnification.

        Although the supervisory signal/information, i.e., ground truth, was weak in the slide-level classification tasks, CAMP was able to attend to pathologically important regions for diagnosis. In other words, without any pixel- and patch-level annotations, the model identified and used critical regions in the slides for diagnosis. 
        For example, for a grade 4 cancer WSI in prostate cancer grading (Fig. \ref{fig_heatmap}a), CAMP clearly attended to malignant tumors, and the highly attended regions showed grade 4 patterns. At high magnification, we observed that CAMP focused on the cribriform pattern of the tumors and ignored loose collagenous stromal tissue.
        In the case of papillary renal cell carcinoma WSI in renal cell cancer sub-typing (Fig. \ref{fig_heatmap}b), CAMP highly focused on the malignant kidney tumors and moderately attended to chronic inflammation and inflammatory areas around the inflamed renal cortical tissue. At the patch level, the glomeruloid growth pattern of the tumor received high attention, while the fibrotic stromal tissue was weakly focused.
        As for breast metastasis detection (Fig. \ref{fig_heatmap}c), all highly attended regions indicated metastases. These regions were surrounded by normal stroma with low attention. Comparing the patches with the high and low attention, we found that the highly attended patches involve malignant regions with solid clusters of tumor cells, whereas the low attention patches show mature small lymphocytes.
        In regard to breast cancer sub-typing (Fig. \ref{fig_heatmap}d), though CAMP highly highlighted the malignant tumors, the predicted label (\textit{ductal carcinoma in situ}) was different from the ground truth label (\textit{invasive carcinoma}). The examination of the highlighted regions provided insight into the wrong classification. Overall, the three regions with high attention showed the pattern of ductal carcinoma in situ. At high magnification, the high-attention patches demonstrated carcinoma with a clinging pattern, whereas loosely fibrotic collagenous stroma was observed in the low-attention patches.

        These observations suggest that CAMP, with attention to heatmaps, permits the interpretation and explanation of the classification results without fine-grained annotations. The ability to recognize essential pathology regions, e.g. tumors, is particularly useful for generating pseudo-labels since the annotation process is time-consuming and labor-intensive. However, we note that the specific meanings of the (highlighted) regions may vary depending on the level of attention, type of WSIs and tasks, and other factors. The detailed interpretation of the results still requires a manual inspection by experienced human experts.

\section{Discussion}
    Computational pathology, powered by advanced AI techniques, has facilitated automated and precise analysis and diagnosis of pathology images. The adoption of computational pathology holds excellent potential for significantly transforming and easing the workflow of conventional pathology. 
    Image classification accounts for a large proportion of pathology tasks. For this reason, a vast amount of research effort in computational pathology has been made to improve the accuracy and reliability of the classification tasks. However, traditional computational pathology approaches do not consider efficiency and scalability with respect to computational costs and resources. In this study, we demonstrated that CAMP is the solution for image classification tasks in pathology that achieves accuracy, reliability, efficiency, and scalability.

    Most of the previous studies in pathology image classification focused on a specific disease, including a single dataset or, at most, a few datasets. The applicability and adaptability of these methods were independently and individually assessed, i.e., task-specific. Though these models have been successfully applied to several tasks in computational pathology and other domains, there has been, to the best of our knowledge, no such study that sought to validate and test over 22 datasets from 17 pathology patch- and slide-level classification tasks. The experimental results in this study showed that the performance of these models considerably varies across different tasks and datasets, questioning the diagnostic accuracy and reliability in the clinics.
    In addition, CAMP is a versatile framework that can handle both patch- and slide-level classification tasks. The former allows CAMP to be employed for categorizing fine-grained pathological characteristics in the region-of-interest level, whereas the latter facilitates the diagnosis at the course-grained slide-level. 
    Moreover, in the previous studies, the efficiency and scalability of these models were not considered in regard to the number of tasks in pathology. With the growing interest and concern in computational resources, these issues need to be taken into account at the developmental stage of computational pathology tools to transform and reshape the current pathology workflow and realize computational pathology in practice.
    Based upon the classification results and the analyses of the computation time and memory consumption, CAMP exhibited the potential for addressing the current and emerging issues and for improving diagnostic accuracy and reliability in pathology.

    This study has several limitations. 
    First, although CAMP is able to adapt to a new task in an efficient and effective fashion, it is not designed to adapt to a new task without annotated examples, i.e. zero-shot learning. Previous models, such as PLIP, were shown to be capable of conducting zero-shot image classification; however, one needs to provide the appropriate prompts and the performance is not only sub-optimal compared to other learning paradigms but also dependent on the quality of prompts \cite{zhou2022prompt}, thereby reducing the chance of the routine use in the clinics. 
    Second, CAMP shares the existing common knowledge but independently and individually learns the task-specific knowledge for the downstream classification tasks. The knowledge learned from the downstream tasks is not shared among the other tasks or used to advance the common knowledge. Ensemble or federated learning approaches could be explored to aggregate the task-specific knowledge and to update the common knowledge without loss of generality. Our future work will investigate the mechanism that can harmonize the existing common knowledge and the new knowledge from various downstream tasks.
    Third, CAMP was successfully applied to 17 classification tasks on both patch- and slide-level, of which 3 tasks included external, independent test datasets. It is generally accepted that the performance could vary on such test datasets due to several reasons, such as variations in slide preparation and image quality \cite{hossain2018practical,hashimoto2012referenceless}. For the tasks with the independent test datasets, CAMP was still the best-performing model compared to other competitors. For the rest of the classification tasks, an additional validation study needs to be followed to verify the superiority of CAMP.
    Fourth, we examined the performance of CAMP using 22 datasets; however, most of the datasets are related to cancer diagnosis. There exist numerous types of classification tasks in computational pathology, such as artifacts detection \cite{histoqc}, survival prediction \cite{zhu2017wsisa,fuchs2008computational}, and treatment response prediction \cite{ali2016computational,meti2021machine}. CAMP is a generic and general framework that can conduct such classification tasks without modifications in the model design.

    With superior performance across extensive and diverse classification tasks, CAMP represents a fundamental transformation in the field of computational pathology for image classification tasks. It moves away from the traditional discriminating methods towards generative techniques, shifts from the category assignment to the production of textual descriptions, and evolves from the static learning to the dynamic and continuous learning approach. We anticipate that CAMP can serve as a universal framework for any classification tasks in pathology, paving the way for the fully digitized and computerized practice of pathology.

\backmatter

\bmhead{Data availability}
Colon-1, Colon-2, UHU, UBC, Gastric, K19, K16, BACH, UniToPatho, PCam, BRACS, and HunCRC are publicly available and can be accessed from the following: Colon-1 and Colon-2 (\url{https://github.com/QuIIL/KBSMC_colon_cancer_grading_dataset}), UHU (\url{https://dataverse.harvard.edu/dataset.xhtml?persistentId=doi:10.7910/DVN/OCYCMP}), UBC (\url{https://gleason2019.grand-challenge.org}),  Gastric (\url{https://github.com/QuIIL/KBSMC_gastric_cancer_grading_dataset}), K19 and K16 (\url{https://zenodo.org/record/53169}), BACH (\url{https://zenodo.org/records/3632035}), UniToPatho (\url{https://zenodo.org/record/4643645}), PCam (\url{https://github.com/basveeling/pcam}), BRACS (\url{https://www.bracs.icar.cnr.it/}), HunCRC$_{P}$ (\url{https://doi.org/10.6084/m9.figshare.c.5927795.v1}), and HunCRC$_{W}$ (\url{https://doi.org/10.7937/tcia.9cjf-0127}).
AGGC, WSSS4LUAD, PANDA, CAMELYON16 are the challenge data that can be accessed at AGGC (\url{https://aggc22.grand-challenge.org}), WSSS4LUAD (\url{https://wsss4luad.grand-challenge.org/WSSS4LUAD}), PANDA (\url{https://panda.grand-challenge.org/home/}) and CAMELYON16 (\url{https://camelyon16.grand-challenge.org/}). 
For KMC-Liver, KMC-Kidney, Bladder, and DHMC, data access shall be addressed to the corresponding authors: KMC-Liver (\url{(https://link.springer.com/article/10.1007/s11042-023-15176-5}), KMC-Kidney (\url{https://github.com/shyamfec/RCCGNet}), Bladder (\url{https://figshare.com/articles/dataset/Bladder_Whole_Slide_Dataset/8116043}), and DHMC (\url{https://bmirds.github.io/KidneyCancer/})

\bmhead{Code availability}
All the details of code/packages and implementation are available at https://github.com/QuIIL/CAMP. All the experiments were run in Python 3.9 with torch v2.0.0, openCV v4.8.1.78, CUDA v11.7.99. Additional packages include tensorboard (2.12.1),  torchvision (0.15.1), timm(0.5.4), grad-cam (1.4.6). All figures were drawn in Microsoft PowerPoint and seaborn v0.13.0. Pretrained models were obtained from open sources and previous works: PLIP (\url{https://huggingface.co/vinid/plip}), CTransPath (\url{https://github.com/Xiyue-Wang/TransPath}), GPC (\url{https://github.com/QuIIL/GPC}), GIT-B (\url{https://huggingface.co/docs/transformers/en/model_doc/git}), UNI (\url{https://huggingface.co/MahmoodLab/UNI}), Phikon (\url{https://huggingface.co/owkin/phikon}), and other models (\url{https://pytorch.org/vision/stable/models.html}).

\bmhead{Acknowledgments}
We acknowledge the support of the National Research Foundation of Korea (NRF) (No. 2021R1A2C2014557) and Institute of Information \& communication Technology Planning \& evaluation (IITP) (No. RS-2022-00167143), funded by the Korea goverment (MSIT).

\bmhead{Author contributions}

J.T.K. conceived, designed, and supervised the study. A.T.N. and J.T.K. performed model design and validation and prepared the manuscript.
A.T.N. developed and tested the Python code and packages and performed experiments. 
A.T.N., K.K., B.S., S.C., and J.T.K. performed data analysis.  
K.K., B.S., and S.C. provided knowledge support with interpreting the results and findings and helped with manuscript preparation. 
All authors contributed to writing the manuscript and reviewed and approved the final version.

\bmhead{Competing interests}
The authors declare no competing interests.

\bibliography{ref}%

\end{document}


\begin{appendices}
\section*{Supplementary Figures}

    \begin{figure*}[!ht]%
        \centering
        \includegraphics[width=1.0\textwidth]{matrix_all_2.png}
        \caption{Confusion matrices of CAMP on 15 datasets.}\label{confusion_matrices_appendix}
    \end{figure*}

    \begin{figure*}[!ht]%
        \centering
        \includegraphics[width=1.0\textwidth]{figures/results_appendix_small.png}
        \caption{Results of 10 classification tasks on 15 datasets by CAMP and 14 competitors. ($\bold{a}$) A radar chart and ($\bold{b}$) bar plots show F1 of CAMP and 14 competitors. 14 competitors include 10 task-specific models and 4 task-agnostic models. 10 task-specific models are 5 CNN models (ConvNeXt-B, EfficientNetV2-S, ResNet50, RegNet, and ResNeXt50) and 5 Transformer models (MaxViT, SwinV2-B, ViT-B, PLIP-V, and CTransPath). 4 task-agnostic models are 2 generative models (GPC and GIT-B), 1 CNN model (ConvNeXt-B$_{TA}$), and 1 Transformer model (CTransPath-B$_{TA}$).}\label{results_appendix}
    \end{figure*}

\section*{Supplementary Table}

    \begin{sidewaystable}[!ht]
    \caption{Performance of CAMP and 14 competitors on 15 datasets.}\label{full_table}
    \fontsize{6.5pt}{6.5pt}\selectfont
    \begin{tabular*}{0.94\textwidth}{p{0.09\textwidth}
    p{0.02\textwidth}p{0.02\textwidth}p{0.02\textwidth}p{0.02\textwidth}
    p{0.00001\textwidth}p{0.02\textwidth}p{0.02\textwidth}p{0.02\textwidth}p{0.02\textwidth}
    p{0.00001\textwidth}p{0.02\textwidth}p{0.02\textwidth}p{0.02\textwidth}p{0.02\textwidth}
    p{0.00001\textwidth}p{0.02\textwidth}p{0.02\textwidth}p{0.02\textwidth}p{0.02\textwidth}
    p{0.00001\textwidth}p{0.02\textwidth}p{0.02\textwidth}p{0.02\textwidth}p{0.02\textwidth}}
    
    \toprule%
    & \multicolumn{4}{@{}c@{}}{Colon-1} && \multicolumn{4}{@{}c@{}}{Colon-2} && \multicolumn{4}{@{}c@{}}{Gastric} && \multicolumn{4}{@{}c@{}}{UHU}  && \multicolumn{4}{@{}c@{}}{UBC}\\ 
    \cmidrule{2-5}\cmidrule{7-10}\cmidrule{12-15}\cmidrule{17-20}\cmidrule{22-25}%
    Model & $Acc$ (\%) & $Acc_c$ (\%)  &  $F1$   & $K_w$ & & 
            $Acc$ (\%) & $Acc_c$ (\%)  &  $F1$   & $K_w$ & & 
            $Acc$ (\%) & $Acc_c$ (\%)  &  $F1$   & $K_w$ & &
            $Acc$ (\%) & $Acc_c$ (\%)  &  $F1$   & $K_w$ & &
            $Acc$ (\%) & $Acc_c$ (\%)  &  $F1$   & $K_w$\\
    \midrule
    ConvNeXt-B   & 
              88.5  & 83.9 & 0.825 & 0.945 & & 
               76.2  & 72.5  & 0.693  &  0.823 & &
               82.3 & 67.8  & 0.739 & 0.897  & & 
               68.7 & 69.2 & 0.592  & 0.738 & & 
               74.8  & 78.2  & 0.622 & 0.683 \\
    EfficientNetV2-S  &  
                86.0  & 80.7  & 0.829  & 0.808  & & 		
                76.9  & 68.9  & 0.699  & 0.701 & &
                81.3 & 68.1  & 0.721 & 0.890  &  &
                69.7 & 66.4 & 0.600  & 0.504 & &
                74.3  & 75.7  & 0.633 & 0.683  \\
    ResNet50  & 
                87.0  & 82.1  & 0.819  & 0.937  & & 
                79.5  & 68.2  & 0.712  & 0.688 & &
                82.2 & 66.9  & 0.722 & 0.901  &  &
                70.9 & 67.5 & 0.599  & 0.512 & &
                77.3  & 78.7  & 0.615 & 0.619\\
    RegNet  & 
                87.9  & 83.2  & 0.831  & 0.943  & &
                82.3  & 76.8  & 0.719  & 0.864 & &
                82.8 & 67.1  & 0.751 & 0.927  &  & 	
                71.2 & 72.2 & 0.611  & 0.639 & &
                76.3  & 77.2  & 0.639 & 0.685\\
    ResNeXt50  & 
                82.4  & 66.4  & 0.828  & 0.918  & &
                79.5  & 67.9  & 0.707  & 0.846 & &
                80.7 & 69.8  & 0.727 & 0.902  &  &
                70.3 & 70.5 & 0.613  & 0.598 & &
                75.9  & 78.6  & 0.624 & 0.695\\
    CTransPath   & 
                88.4  & 83.7 & 0.847 & 0.945 & & 
               81.3  & 75.6  & 0.743  &  0.898 & &   	
               84.6 & 70.6  & 0.765 & 0.934   & &  	
               69.3 & 70.2 & 0.615  & 0.614 & & 
               76.6  & 76.8  & 0.663 & 0.691 \\ 
    PLIP-V  &  
                84.4  & 78.3  & 0.815  & 0.924  & & 		
                72.9  & 65.4  & 0.711  & 0.831 & & 
                79.7 & 75.0  & 0.739 & 0.861  &  &
                67.8 & 66.2 & 0.573  & 0.587 & &
                73.6  & 74.1  & 0.627 & 0.675\\
    ViT-B  & 
                87.5  & 82.0 & 0.833  & 0.838  & &
                79.8  & 72.8  & 0.702  & 0.899 & &
                84.4 & 69.2  & 0.724 & 0.930 &  & 
                71.9 & 72.2 & 0.609  & 0.643 & &       
                75.4  & 75.9  & 0.605 & 0.690\\
    SwinV2-B  &
                86.5  & 81.2  & 0.822  & 0.933  & & 
                70.4  & 69.0  & 0.671  & 0.842 & &
                83.9 & 68.5  & 0.774 & 0.935  &  &  
                71.9 & 72.0 & 0.606  & 0.639 & &       
                73.9  & 75.1  & 0.598 & 0.669\\
    MaxViT  &
                87.9  & 84.0  & 0.829  & 0.805  & &
                76.3  & 72.8  & 0.689  & 0.895 & & 
                83.2 & 68.5  & 0.730 & 0.926  &  &
                71.6 & 70.2 & 0.597  & 0.649 & &  
                75.9  & 76.7  & 0.631 & 0.678\\    
    ConvNeXt-B-TA  &
                82.4  & 75.6  & 0.811  & 0.784  & &
                74.2  & 72.7  & 0.729  & 0.874 & &
                81.8 & 66.4  & 0.758 & 0.854  &  &
                68.3 & 66.8 & 0.610  & 0.602 & &
                73.3 & 74.2  & 0.645 & 0.638\\
    CTransPath-TA  &
                82.5  & 81.2  & 0.812  & 0.797  & &
                75.3  & 74.2  & 0.733  & 0.853 & &
                82.7 & 69.4   & 0.763 & 0.844  &  &
                69.3 & 67.7 & 0.604  & 0.593 & &
                74.3 & 75.3  & 0.663 & 0.642\\
    GIT-B  & 
                85.3  & 80.4  & 0.831  & 0.913  & &  
                70.8  & 69.7  & 0.711  & 0.858 & & 
                83.2 & 67.9  & 0.731 & 0.897  &  &
                68.8 & 69.4 & 0.606  & 0.619 & &
                72.9  & 74.8  & 0.657 & 0.691\\
    GPC  & 
                88.4  & 83.8  & 0.848  & 0.944  & &
                79.0  & 74.0  & 0.722  & 0.898 & &  
                83.7 & 83.7  & 0.768 & 0.925  &  & 
                70.4 & 71.9 & 0.628  & 0.612 & & 
                76.9 & 79.0  & 0.641 & 0.700\\
    CAMP (ours) &
                89.2  & 84.9  & 0.859  & 0.949  & & 
                84.0  & 78.8  & 0.775  & 0.914 & & 	
                85.2 & 71.5  & 0.788 & 0.929  &  &  			
                71.2 & 72.2 & 0.629  & 0.630 & &
                79.9  & 79.8  & 0.715 & 0.725\\\\
    \midrule
    & \multicolumn{4}{@{}c@{}}{AGGC} && \multicolumn{4}{@{}c@{}}{PANDA} && \multicolumn{4}{@{}c@{}}{K19} && \multicolumn{4}{@{}c@{}}{K16}  && \multicolumn{4}{@{}c@{}}{KMC-Liver}\\ 
    \cmidrule{2-5}\cmidrule{7-10}\cmidrule{12-15}\cmidrule{17-20}\cmidrule{22-25}%
    Model & $Acc$ (\%) & $Acc_c$ (\%)  &  $F1$   & $K_w$ & & 
            $Acc$ (\%) & $Acc_c$ (\%)  &  $F1$   & $K_w$ & & 
            $Acc$ (\%) & $Pre$ (\%)  &  $F1$   & $K_w$ & &
            $Acc$ (\%) & $Pre$ (\%)  &  $F1$   & $K_w$ & &
            $Acc$ (\%) & $Acc_c$ (\%)  &  $F1$   & $K_w$\\
    \midrule
    ConvNeXt-B   & 
               56.8  & 63.3 & 0.410 & 0.524 & & 
               91.2  & 90.6  & 0.855  & 0.897 & &  	
               99.7 & 0.995  & 0.995 & 0.997  & & 
               73.9 & 0.754 & 0.731  & 0.778 & &  
               91.4  & 88.5  & 0.922 & 0.940 \\
    EfficientNetV2-S  &  
                55.4  & 59.9  & 0.398  & 0.505  & &
                89.5  & 88.7  & 0.863  & 0.894 & & 			
                99.7 & 0.996  & 0.996 & 0.996  &  &
                53.3 & 0.634 & 0.747  & 0.488 & & 
                95.7  & 94.3  & 0.936 & 0.980\\
    ResNet50  & 
                56.3  & 61.5  & 0.427  & 0.531  & & 
                90.4  & 90.7  & 0.834  & 0.900 & &  
                99.7 & 0.996  & 0.996 & 0.996  &  &
                48.8 & 0.601 & 0.749  & 0.488 & &
                94.6  & 92.9  & 0.940 & 0.975\\
    RegNet  & 
                56.6  & 62.6  & 0.432  & 0.551  & &
                90.7  & 90.9  & 0.870  & 0.906 & & 		
                99.7 & 0.997  & 0.997 & 0.997  &  &
                67.1 & 0.646 & 0.754  & 0.671 & &
                95.0  & 93.3  & 0.952 & 0.967\\
    ResNeXt50  & 
                54.7  & 60.8  & 0.428  & 0.548  & &
                90.2  & 90.1  & 0.851  & 0.897 & & 		
                99.6 & 0.996  & 0.995 & 0.995  &  &
                53.3 & 0.817 & 0.733  & 0.533 & & 
                94.6  & 92.8  & 0.914 & 0.967\\
    CTransPath   & 
                56.7  & 61.5 & 0.427 & 0.552 & &  			
               91.5  & 91.7  & 0.860  &  0.905 & & 
               99.6 & 0.996  & 0.997 & 0.996  & &  
               61.0 & 0.694 & 0.809  & 0.610 & &
               95.7  & 94.3  & 0.962 & 0.980 \\
    PLIP-V  &  
                56.6  & 59.7  & 0.400  & 0.496  & &
                88.9  & 89.5  & 0.842  & 0.878 & &
                99.5 & 0.993  & 0.993 & 0.993  &  &
                67.1 & 0.813 & 0.686  & 0.671 & &  
                93.6  & 94.5  & 0.941 & 0.964\\
    ViT-B  & 
                55.9  & 58.2  & 0.434  & 0.523  & &
                90.2  & 90.9  & 0.871  & 0.881 & &
                99.6 & 0.995  & 0.995 & 0.995  &  &
                59.5 & 0.752 & 0.728  & 0.595 & & 
                95.7  & 93.3  & 0.938 & 0.937\\
    SwinV2-B  &
                57.6  & 59.3  & 0.462  & 0.547  & &
                90.9  & 91.2  & 0.883  & 0.908 & & 	
                99.7 & 0.997  & 0.997 & 0.997  &  &
                72.2 & 0.712 & 0.804  & 0.722 & & 
                94.6  & 92.8  & 0.959 & 0.971\\
    MaxViT  &
                55.0  & 57.7  & 0.429  & 0.539  & &
                90.7  & 90.8  & 0.881  & 0.899 & &	
                99.5 & 0.995  & 0.994 & 0.995  &  & 
                64.3 & 0.815 & 0.744  & 0.643 & & 
                95.4  & 95.2  & 0.918 & 0.972 \\
    ConvNeXt-B-TA  &
                55.6  & 58.2  & 0.421  & 0.521  & &
                89.3  & 90.3  & 0.843  & 0.893 & &
                99.2 & 0.992  & 0.985 & 0.989  &  &
                52.6 & 0.684 & 0.799  & 0.524 & &
                91.8  & 90.3  & 0.949 & 0.923\\
    CTransPath-TA  &
                56.1  & 58.7  & 0.431  & 0.528  & &
                88.8  & 89.7  & 0.851  & 0.906 & &
                99.4 & 0.993  & 0.989 & 0.991  &  &
                57.1 & 0.690 & 0.801  & 0.711 & &
                92.5  & 93.6  & 0.938 & 0.943\\
    
    GIT-B  & 
                56.6  & 62.1  & 0.412  & 0.553  & &
                88.4  & 87.6  & 0.832  & 0.835 & &
                99.4 & 0.995  & 0.993 & 0.993  &  &
                54.9 & 0.471 & 0.709  & 0.582 & &
                92.3  & 94.5  & 0.931 & 0.944\\
    GPC  & 
                57.4  & 63.4  & 0.459  & 0.522  & &
                88.9  & 90.7  & 0.848  & 0.904 & &
                99.5 & 0.995  & 0.995 & 0.995  &  &
                56.9 & 0.532 & 0.787  & 0.649 & &
                94.2  & 93.1  & 0.950 & 0.955 \\
    CAMP (ours) &
                60.8  & 64.8  & 0.488  & 0.575  & &  	
                91.8  & 92.0  & 0.893  & 0.915 & & 	
                99.8 & 0.998  & 0.998 & 0.998  &  &
                76.0 & 0.736 & 0.813  & 0.760 & &
                96.1  & 96.1  & 0.962 & 0.982\\\\
    \midrule
    & \multicolumn{4}{@{}c@{}}{KMC-Kidney} && \multicolumn{4}{@{}c@{}}{Bladder} && \multicolumn{4}{@{}c@{}}{BACH} && \multicolumn{4}{@{}c@{}}{PCam}  && \multicolumn{4}{@{}c@{}}{WSSS4LUAD}\\ 
    \cmidrule{2-5}\cmidrule{7-10}\cmidrule{12-15}\cmidrule{17-20}\cmidrule{22-25}%
    Model & $Acc$ (\%) & $Acc_c$ (\%)  &  $F1$   & $K_w$ & & 
            $Acc$ (\%) & $Acc_c$ (\%) & $F1$  & $Re$  & & 
            $Acc$ (\%) & $Acc_c$ (\%) & $F1$  & $Re$ & &
            $Acc$ (\%) & $Pre$  & $F1$ & $Re$ & &
            $Acc$ (\%) & $Pre$  & $F1$ & $Re$ \\
    \midrule
    ConvNeXt-B   & 
                88.0 & 84.1 & 0.842 & 0.972 & & 
                90.1 & 80.1  & 0.803 & 0.838  & & 
               78.4 & 64.2 & 0.763  & 0.751 & & 
               89.7 & 0.880 & 0.856  & 0.879 & & 
               92.1  & 82.4  & 0.805 & 0.876 \\
    EfficientNetV2-S  &  
                88.0  & 88.2  & 0.858  & 0.962  & &
                89.7  & 88.5  & 0.788  & 0.897 & &
                80.2 & 62.5  & 0.759 & 0.762  &  &
                89.5 & 0.871 & 0.841  & 0.824 & &
                89.7 & 80.3  & 0.811 & 0.902\\
    ResNet50  & 
                88.0  & 85.1  & 0.871  & 0.945  & &
                89.8  & 76.8  & 0.794  & 0.837 & &
                74.0 & 62.7  & 0.736 & 0.708  &  &
                88.9 & 0.888 & 0.947  & 0.889 & &
                90.9  & 81.4  & 0.820 & 0.921\\
    RegNet  & 
                89.8  & 90.0  & 0.871  & 0.956  & &
                89.4  & 85.2  & 0.805  & 0.839 & &
                80.2 & 71.4  & 0.782 & 0.811  &  &
                88.9 & 0.890 & 0.982  & 0.891 & &
                91.1  & 81.6  & 0.820 & 0.935\\
    ResNeXt50  & 
                89.4  & 86.1  & 0.877  & 0.947  & &
                89.1  & 79.9  & 0.801  & 0.814 & &
                75.4 & 66.8  & 0.775 & 0.774  &  &
                88.3 & 0.891 & 0.891  & 0.872 & &
                88.7  & 80.1  & 0.809 & 0.928\\
    CTransPath   & 
                90.8  & 90.1 & 0.901 & 0.961 & & 
               90.8  & 90.1  & 0.891  &  0.961 & & 
               80.5 & 71.4  & 0.722 & 0.822  & &  
               88.6 & 0.897 & 0.980  & 0.886 & & 
               92.1  & 81.4  & 0.836 & 0.947 \\ 
    PLIP-V  &  
                88.7  & 85.1  & 0.833  & 0.965  & &
                87.9  & 78.2  & 0.866  & 0.823 & &
                76.3 & 61.2  & 0.627 & 0.821  &  &
                88.4 & 0.885 & 0.895  & 0.854 & &
                90.4  & 80.7  & 0.803 & 0.898\\
    ViT-B  & 
                86.6  & 84.1  & 0.825  & 0.945  & &
                87.1  & 82.2  & 0.856  & 0.833 & &
                78.3 & 72.5  & 0.734 & 0.766  &  &
                87.2 & 0.885 & 0.953  & 0.862 & &
                90.9  & 81.8  & 0.827 & 0.916\\
    SwinV2-B  &
                89.5  & 88.3  & 0.889  & 0.951  & &
                89.8  & 80.1  & 0.879  & 0.834 & &
                78.9 & 74.5  & 0.743 & 0.753  &  &
                88.1 & 0.899 & 0.880  & 0.891 & &
                90.4  & 80.9  & 0.980 & 0.906\\
    MaxViT  &
                90.1  & 88.1  & 0.814  & 0.953  & &
                90.4  & 76.4  & 0.882  & 0.845 & &
                78.9 & 71.4  & 0.698 & 0.742  &  &
                88.7 & 0.876 & 0.863  & 0.877 & &
                89.8  & 80.6  & 0.816 & 0.928\\
    ConvNeXt-B-TA  &
                84.5  & 85.1  & 0.870  & 0.945  & &
                87.4  & 78.8  & 0.869  & 0.818 & &
                77.4 & 62.7  & 0.761 & 0.764  &  &
                85.8 & 0.859 & 0.965  & 0.855 & &
                89.5  & 80.4  & 0.811 & 0.938\\
    CTransPath-TA  &
                85.2  & 84.1  & 0.865  & 0.947  & &
                88.5  & 87.4  & 0.851  & 0.922 & &
                80.2 & 65.8  & 0.769 & 0.734  &  &
                88.6 & 0.886 & 0.970  & 0.886 & &
                89.9  & 80.9  & 0.838 & 0.921\\
    GIT-B  & 
                81.7  & 86.1  & 0.836  & 0.965  & &
                88.1  & 83.8  & 0.830  & 0.925 & &
                80.7 & 69.9  & 0.726 & 0.780  &  &
                86.0 & 0.789 & 0.929  & 0.775 & &
                91.5  & 81.6  & 0.798 & 0.919\\
    GPC  & 
                87.3  & 88.3  & 0.888  & 0.972  & &
                87.3  & 87.4  & 0.871  & 0.942 & &
                83.4 & 65.3  & 0.779 & 0.728  &  &
                87.3 & 0.864 & 0.975  & 0.792 & &
                91.0  & 81.4  & 0.819 & 0.890\\
    CAMP (ours) &
                90.8  & 90.8  & 0.890  & 0.967  & & 	
                90.8  & 90.8  & 0.902  & 0.967 & &
                83.9 & 80.1  & 0.791 & 0.825  &  &
                88.8 & 0.900 & 0.982 & 0.888 & & 		
                93.6  & 83.8  & 0.798 & 0.947\\
    
    \botrule
    \end{tabular*}
\end{sidewaystable}

\end{appendices}